\documentclass{siamltex}
\usepackage{amsmath}
\usepackage{amssymb}
\usepackage{dsfont}
\usepackage{color}
\usepackage{graphicx}
\usepackage{subfig}
\usepackage{multirow}
\usepackage{morefloats}
\usepackage{pgfplots}
\usetikzlibrary{plotmarks}
\usepackage{gensymb}
\usepackage{tikz}

\begin{document}
\title{Simulation of Fluid Particle Cutting - Validation and Case Study}

\author{M. W. Hlawitschka\footnotemark[1] \and S. Tiwari \footnotemark[2]  \and J. Kwizera\footnotemark[1]
\and  H.-J. Bart \footnotemark[1] \and A. Klar \footnotemark[2] \footnotemark[3]}
 \footnotetext[1]{Technische Universit\"at Kaiserslautern, Department of Mechanical and Process Engineering, Erwin-Schr\"odinger-Stra{\ss}e, 67663 Kaiserslautern, Germany 
    (\{bart,hlawitschka\}@mv.uni-kl.de, kwizera@rhrk.uni-kl.de)}
    \footnotetext[2]{Technische Universit\"at Kaiserslautern, Department of Mathematics, Erwin-Schr\"odinger-Stra{\ss}e, 67663 Kaiserslautern, Germany 
      (\{klar,tiwari\}@mathematik.uni-kl.de)}
\footnotetext[3]{Fraunhofer ITWM, Fraunhoferplatz 1, 67663 Kaiserslautern, Germany}

%\author{R. Borsche\footnotemark[1] 
%     \and  A. Klar\footnotemark[1] \footnotemark[2]}
%\footnotetext[1]{Technische Universit\"at Kaiserslautern, Department of Mathematics, Erwin-Schr\"odinger-Stra{\ss}e, 67663 Kaiserslautern, Germany 
%  (\{borsche, klar\}@mathematik.uni-kl.de)}
%\footnotetext[2]{Fraunhofer ITWM, Fraunhoferplatz 1, 67663 Kaiserslautern, Germany} 
% 
\date{}

\maketitle
\begin{abstract}
\noindent  In this paper we  present the comparison of experiments and numerical simulations for bubble cutting by a wire. The air bubble is surrounded by water. In the experimental setup an air bubble is injected on the bottom of a water column. When the bubble rises and contacts the wire, it is separated into two daughter bubbles. 
The flow is modeled by the incompressible Navier-Stokes equations. A meshfree method is used to simulate the bubble cutting.  We have observed that the experimental and numerical results are in very good agreement.  Moreover, we have further presented simulation results for  liquid with higher viscosity. In this case the numerical results are close to previously published results. 
\end{abstract}


\section{Introduction}
  Fluid particle cutting plays an important role in gas-liquid and  liquid-gas contactors. In gas-liquid contactors, the bubble size distribution, determining the mass transfer area, is influenced by the local hydrodynamics, but also by measuring probes such as needle probes \cite{Mudde, Ahmed, Choi} and mesh based conductivity probes \cite{Prasser}. The shape of the probes are mainly cylindrical, while the probe may be in flow direction but also in a rectangular angle to it.  The rising bubbles approach the immersed object and starts to change its shape. Depending on the position of the bubble to the wire, the bubble will pass the object or be cutted in two fragments (daughter bubbles). Beside the unwanted cutting at probes, a wire mesh can be used to generate smaller bubbles and homogenise the flow structure. \\
Furthermore, in liquid-gas contactors, phase separation is often a problem. Demisters are then frequently used to prevent a phase slip (entrainment) of fine dispersed phase droplets in the continuous product phase.  A loss of the total solvent inventory within one year is reported causing costs and environmental hazards. Entrainment can cause a significant reduction in separation efficiency. Demisters are based on wire meshes, where the small droplets should accumulate. Using  an optimal design, the small droplet seperation efficiencies can be up to 99.9\%. Nevertheless, bigger droplets tend to break up in the rows of wires.

Hence, particle cutting is a frequently observed phenomena in various separation processes ranging from low viscosity to high viscosity of the continuous fluid. Nevertheless, it can be hardly investigated under operation conditions due to the complex insertion of optical probes into the apparatus or the complex mesh structure e.g. of the demister, but also the operation conditions as high pressure, high dispersed phase hold ups make an experimental investigation challenging. 

In this study, we focus on the simulation of particle cutting at a single wire strengthened by experimental investigations to generate the basis for further numerical studies at complex geometries and fluid flow conditions such as demister simulations.
For the simulation of bubble cutting, a meshfree approach is applied. It overcomes several drawbacks of classical computational fluid dynamics (CFD) methods such as Finite Element Method (FEM) , Finite Volume Method (FVM). The main drawback of the classical methods (FEM, FVM) is the relatively expensive geometrical mesh/grid required to carry out the numerical computations. The computational costs to generate and maintain the grid becomes particularly high for complex geometries and when the grid moves in time, as in the case of fluid particles with a dynamic interface or in  case where the interface between fluids changes in time. 

For such problems meshfree methods are appropriate. Here, we use a meshfree method, based on the generalized finite difference method, called Finite Pointset Method (FPM).  The two phase flow is modeled by using the continuous surface force (CSF) model \cite{CSF}. 
Each phase is indicated by the color of the respective particles. When particles move, they carry all the information about the flow with them such as their color, density, velocity, etc. The colors, densities and viscosity values of all particles remain constant during the time evolution. The fluid-fluid interface is easily determined with the help of the color function  \cite{Morris}. In \cite{TK07} an implementation of the CSF model within the FPM was presented to simulate surface-tension driven flows.  We have further extended the method to simulate wetting phenomena \cite{TKH16}. 
    
 \section{Experimental Setup}
\begin{figure}	
		\includegraphics[keepaspectratio=true, width=0.9\textwidth]{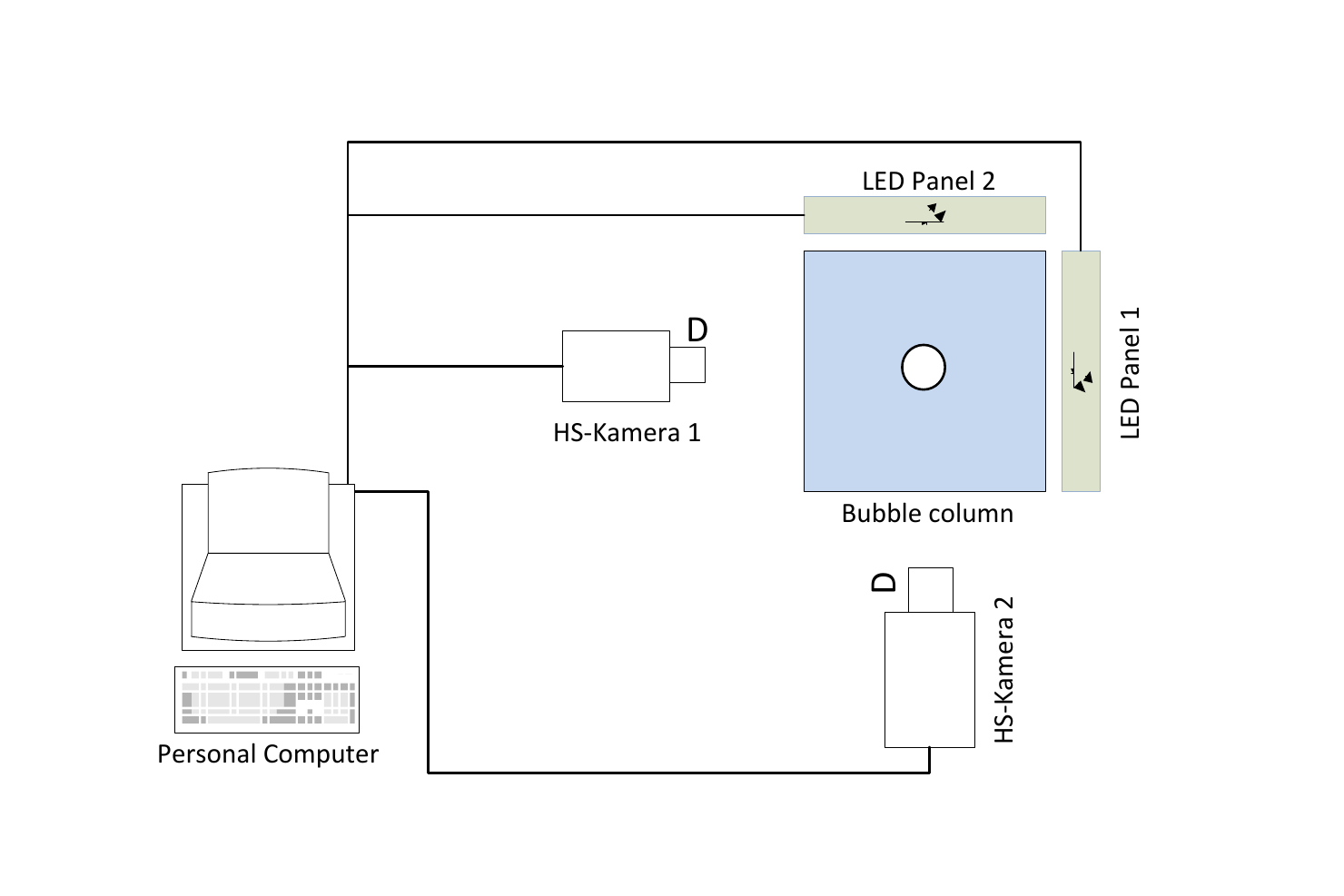}	
	\captionsetup{margin=20pt}
	\caption{Sketch of the experimental setup showing the plexiglas column in the middle (blue).  }
	\label{Sketch}
\end{figure}

Bubble cutting is investigated in a Plexiglas column filled with reversed osmosis water up to a level of 10 cm. The column has a width and depth of 46  mm.  A syringe pump  (PSD/3, Hamilton) is used to inject air of known volume at the bottom of the column. The injection diameter is 8 mm.  The schematic setup is given in Figure \ref{Sketch}. In a distance of 60 mm from the bottom, a wire is mounted  in the middle of the column. The wire has a diameter of 3 mm. Two cameras (Imaging Solutions NX8-S2 and Os 8-S2) are mounted in an angle of 90\degree to track the bubble motion over an image sequence, respectively over time. Both cameras are triggered and allow a synchronous detection at 4000 fps and a resolution of 1600x1200 px$^2$. By tracking the bubble motion from two sides, it is possible to analyse the side movement and to detect the exact position of bubble contact with the wire. Also, the bubble deformation can be analyzed in two direction and therefore leads to more precise results compared to single camera setups.

\section{Bubble Motion Analyses}
\begin{figure}
  \centering
  \begin{tabular}[b]{p{3.5cm}}
    \includegraphics[width=1\linewidth]{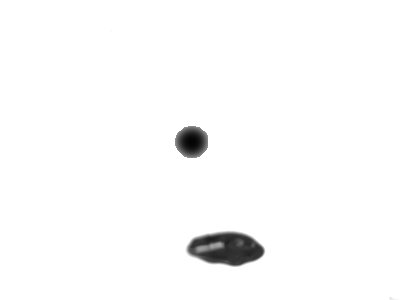} \\
    \small (a) 1. camera raw image.
  \end{tabular} \qquad
  \begin{tabular}[b]{p{3.5cm}}
    \includegraphics[width=1\linewidth]{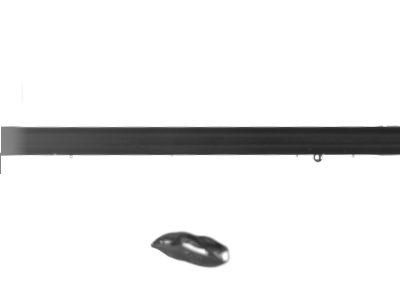} \\
    \small (b) 2. camera raw image.
  \end{tabular} \qquad
  \begin{tabular}[b]{p{5cm}}
    \includegraphics[width=1\linewidth]{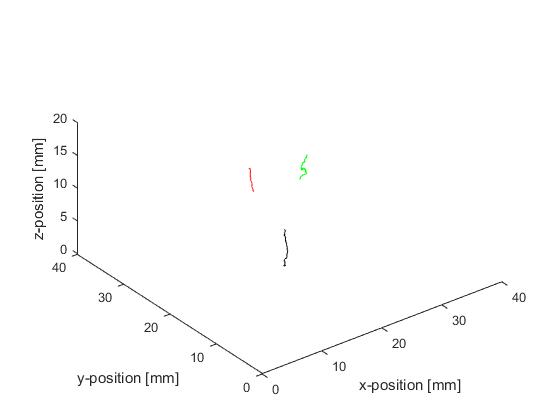} \\
    \small (c) 3d bubble paths.
  \end{tabular}
  \begin{tabular}[b]{p{5cm}}
    \includegraphics[width=1\linewidth]{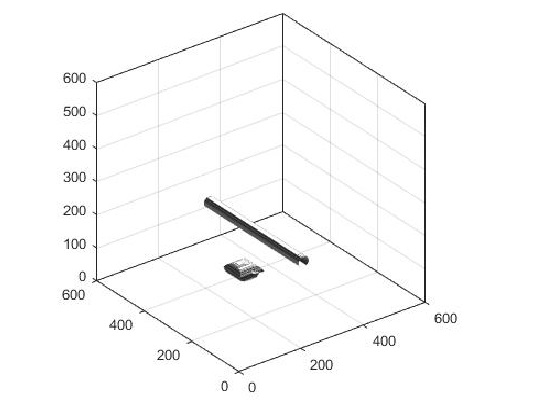} \\
    \small (d) Reconstructed bubble in a three dimensional space.
  \end{tabular}
  \begin{tabular}[b]{p{5cm}}
    \includegraphics[width=1\linewidth]{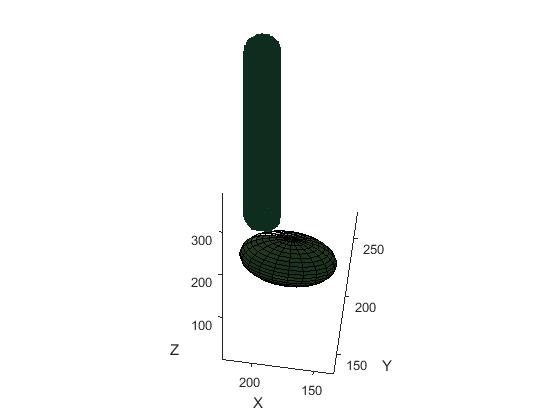} \\
    \small (e) Visualization based on the assumption of ellipsoidic structure.
  \end{tabular}

  \caption{3D bubble reconstruction}
\label{Bubblereconstruction}
\end{figure}

For bubble motion analyses, the tool box ImageJ (https://imagej.nih.gov/ij/) is used. The raw images (Fig. \ref{Bubblereconstruction}  a/b) are therefore binarized, followed by a watershed segmentation. The tracks of the single bubble and cutted particles are analysed using the Plugin Mtrack2 (http://imagej.net/MTrack2). Two particle tracks are tracked, one by each camera and reconstructed using Matlab software toolbox ( (Fig. \ref{Bubblereconstruction}  c) These are the basis for three dimensional reconstruction of the bubble motion. The conversion of pixels to metric length is done by a afore performed calibration. 

Matlab is used to reconstruct the bubble in a three dimensional domain. The images are converted to greyscale and further to binary images. Possible holes (white spots in a surrounded black bubble structure) are filled to get a better identification of the bubbles. To detect the bubble position, a distance transform is performed, followed by a watershed segmentation to separate the bubble from the pipe structure. Finally, the bubble size and shape is determined from each image.
In a next step, the basic grey scale images are again converted to binary images, followed by a watershed algorithm \cite{BMJ}. Finally, the resulting structures are transformed into 3D space. Therefore, the detected structures are extruded into the third dimension resulting in overlapping structures. 
The overlapping structures represent the bubble and the wire and are visualized in Fig. \ref{Bubblereconstruction}d. Applying the assumption of an ellipsoidic structure for the bubble, results finally in  Fig. \ref{Bubblereconstruction}e.

\section{Mathematical Model }
We consider a one-fluid model for two immiscible fluids which are liquid and gas. We model the equation of motion of these fluids by the incompressible Navier-Stokes equations, which are given in the Lagrangian form  
 \begin{eqnarray}
 \frac{d{\bf x}}{dt} &=&{\bf v}\\
\nabla\cdot {\bf v} &= &0 \\
\rho \frac{d{\bf v}}{dt} &= &-\nabla p +  \nabla\cdot(2 \mu \tau) + \rho {\bf g}  +  \rho {\bf F}_S,
\label{ns}
 \end{eqnarray}
where  ${\bf v}$ is the fluid velocity, $\rho$ is the fluid density, $p$ is the pressure, $\tau$ is the stress tensor given by 
$\tau = \frac{1}{2}(\nabla {\bf v} + (\nabla {\bf v})^T)$, $\bf {g}$ is the external force and ${\bf F}_S$ is the surface tension force. The quantity ${\bf F}_S$ is force density, which acts on the interface and its neighbor of the interface between gas and liquid. We compute the surface tension force CSF model of Brackbill {\it et al} (\cite{CSF}) and is 
given by 
{\begin{equation}
{\bf F}_S = \sigma\kappa{\bf n} \delta_S,  
\end{equation}
where, $\sigma$ is the surface tension coefficient, ${\bf n}$ is the unit normal vector at the interface and its neighbor, $\kappa$ is the curvature and $\delta_S$ is the surface delta function. We note that $\delta_S$  is quite strong in the interface and its surroundings.  
We solve the equations (\ref{ns}) with initial and boundary conditions. 

\section{Numerical methods}
We solve the equations (\ref{ns}) by a meshfree Lagrangian particle method. In this method, we first approximate a computational domain by discrete grid points. 
The grids points are divided into two parts as interior and boundary particles.  The boundary particles approximate boundaries and we prescribe boundary conditions on them. 
The interior particles  move with the fluid velocities. Particles may come very  close to each other or can go far away from each other leading to very fine or very coarse approximations.  This problem has to be tackled carefully due to  stability reasons. To obtain a 
uniform distribution of particles in each time step one has to add or remove particles, if  necessary. We refer to \cite{DTKB} for details of such a particle management. 
 
\subsection{Computation of the quantities in surface tension force}
\label{comp_csf}
 For meshfree particle methods  the interfaces between fluids are easily 
tracked by using  flags on the particles. Initially, we assign different flags or color function $c$ of particles representing the corresponding fluids. We define the color function 
$c = 1$ for fluid type 1 and $c=2$ for fluid type 2. On the interface and its vicinity, the Shepard interpolation is
applied for smoothing of the color functions using 
\begin{equation}
\tilde c({\bf x}) = \frac{\sum_{i=1}^{m} w_i c_i}{\sum_{i=1}^{m}},
\end{equation}
where $m$ is the number of neighbor of arbitrary particle having position ${\bf x}$,   $c_i$ are the color values at 
neighboring particle $i$ and $w_i$ is the weight as a function of distance from ${\bf x}$ to $\bf{x}_i$ given by 
\begin{eqnarray}
w_i = w( {\bf x}_i - {\bf x}; h) =
\left\{
\begin{array}{l}
\exp (- \alpha \frac{\| {\bf x}_i - {\bf x}  \|^2 }{h^2} ),
\quad \mbox{if    }  \frac{\| {\bf x}_i - {\bf x}  \|}{h} \le 1
\\  
 0,  \qquad \qquad \quad \quad \quad \quad  \mbox{else}, 
\end{array}
\right.
\label{weight}
\end{eqnarray}
where $ \alpha $ is a positive constant
After smoothing the color function, we compute the unit normal vector 
\begin{equation}
{\bf n} = \frac{ \nabla \tilde c}{ | \nabla \tilde c |}. 
\end{equation}
Finally, we compute the curvature  by 
\begin{equation}
\kappa = -\nabla\cdot{\vec n}. 
\label{curvature}
\end{equation}
The quantity $\delta_s$ is approximated as 
\begin{equation}
\delta_s \approx | \nabla \tilde c |.
\label{deltaS}
\end{equation} 
 Here, $\delta_s$ is non-zero in the vicinity of the interface and vanishes far from it. 

%%%%%%%%%%%%%%%%%%%%%%%%%%%%%%%%%%%%%%%%%%%%%%%%%

\subsection{Numerical scheme}
 \label{num_scheme}
We solve the Navier-Stokes equations (\ref{ns}) with the help of Chorin's projection method \cite{Chorin}. Here, the projection method is adopted in the Lagrangian meshfree 
particle method. Consider the discrete time levels $t^n = n~dt, n = 0,1,2,\ldots $ with time step $dt$. Let ${\bf x}^n$ be the  
position of a particle at time level $n$. 

In the Lagrangian particle scheme we compute the new particle positions at the time level $(n+1)$ by 
\begin{equation}
{\bf x}^{n+1}  =  {\bf x}^n + dt \; {\bf v}^n 
\end{equation}
and then use Chorin's pressure projection scheme in new positions of particles.  
The pressure projection scheme is divided into two steps. 
 The first step consists of computing the  
intermediate velocity $\bf{v}^{*}$ with neglecting the pressure term 
\begin{equation}
{\bf v}^{*} =  {\bf v}^n + \frac{dt}{\rho} \nabla\cdot(2 \mu \tau^{*} )  + dt ~{\bf g} + \frac{dt}{\rho} {\bf F}^n_S. 
\label{intermediatev}
\end{equation}
Since we use the Lagrangian formulation, we do not need to handle the nonlinear convective term.  
The second step consists of computation of pressure and the velocity at time level $(n+1)$  
by solving the equation
\begin{equation}
{\bf v}^{n+1} = {\bf v}^{*} -  dt \; \frac{\nabla p ^{n+1}}{\rho}
\label{correctv}
\end{equation}
where ${\bf v}^{n+1}$ should obey the continuity equation 
\begin{equation}
\nabla\cdot {\bf v}^{n+1} = 0.
\label{constraint}
\end{equation}
We observe from the equation (\ref{correctv}) that the new pressure $p^{n+1}$ is necessary in order 
to compute the new velocity. ${\bf v}^{n+1}$ . Now, we take the divergence of equation (\ref{correctv}) on both sides and use of the continuity constraint (\ref{constraint}), we 
obtain the pressure Poisson equation  
\begin{equation}
\nabla\cdot\left(\frac{ \nabla p^{n+1}}{\rho}\right) = 
\frac{\nabla \cdot {\bf v}^{*}} {dt}.
\label{poisson}
\end{equation}
In order to derive the boundary condition for $p$ we project the equation
(\ref{correctv}) on the outward unit normal vector ${\bf n}$ at the boundary
$\Gamma$ and then we obtain the Neumann boundary condition
\begin{equation}
\left(\frac{\partial p}{\partial {\bf n} }\right)^{n+1} =
- \frac{\rho}{dt} (
{\bf v}^{n+1}_{\Gamma} - {\bf v}^{*}_{\Gamma}) \cdot {\bf n},
\end{equation}
where ${\bf v}_{\Gamma}$ is the value of ${\bf v}$ on $\Gamma$.
Assuming ${\bf v}\cdot {\bf n} = 0$ on $\Gamma$, we obtain
\begin{equation}
\left(\frac{\partial p}{\partial {\bf n} }\right)^{n+1} = 0 
\label{nbc}
\end{equation}
on $\Gamma$.

We note that we have to approximate the spatial derivatives at each particle position 
as well as solve the second order elliptic problems for the velocities and the pressure. 
The spatial derivatives at each particle position are approximated from its neighboring clouds of particles based on the 
weighted least squares method. The weight is a function of a distance of a particle position  to its neighbors. We observe that in 
Eq. \ref{intermediatev} there is a discontinues coefficient $\mu$ inside the divergence operator since the viscosities of two liquid 
may have the ratio of up to 1 to 100. Similarly, the density ratio also has 1 to 1000, which can be seen also in Eq. \ref{poisson}. 
This discontinous coefficients have to be smoothed for stable computation. 
 This is done using a similar procedure as for smoothing the color function. 
We denote the  smoothed viscosity and density by $\tilde\mu$ and $\tilde\rho$, respectively. We note that we smooth the density and viscosity while 
solving Eqs. \ref{intermediatev} and \ref{poisson}, but keep them constant on each phase of particles during the entire computational time. If the density and viscosity has larger ratios, we may have 
to iterate the smoothing  2 or 3 times. 
 Finally,  Eq. \ref{intermediatev} and  
 Eq. \ref{poisson} can be re-expressed as 
\begin{eqnarray}
u^{*} -  \frac{dt}{\tilde\rho} \nabla\tilde\mu\cdot\nabla u^{*} - 
dt \frac{\tilde\mu}{\tilde\rho} \Delta u^{*} 
 &=&  
u^n + dt \; g_x +   
\frac{dt}{\rho}(\frac{\partial\tilde\mu}{\partial x}\frac{\partial u^n}{\partial x} +
\frac{\partial \tilde\mu}{\partial y}\frac{\partial v^n}{\partial x}  )
\label{intu}
\\
v^{*} -  
\frac{dt}{\tilde\rho} \nabla\tilde\mu\cdot\nabla v^{*} - 
dt \frac{\tilde\mu}{\tilde\rho} \Delta v^{*} 
  &= & 
v^n + dt \; g_y +   
\frac{dt}{\tilde\rho}(\frac{\partial\tilde\mu}{\partial x}\frac{\partial u^n}{\partial y} +
\frac{\partial \tilde \mu}{\partial y}\frac{\partial v^n}{\partial y} )
\label{vint}
\\
-\frac{\nabla\tilde\rho}{\tilde\rho}\cdot\nabla p^{n+1} + \Delta p^{n+1} &=& \tilde\rho\frac{\nabla\cdot{\vec v}^{*}}{dt}.
\label{poisson1}
\end{eqnarray}
 Note that, for constant density,  the first term of Eq. \ref{poisson1} vanishes and 
we get the pressure Poisson equation. Far from the interface we have $\tilde\mu = \mu$ and $\tilde\rho = \rho$. 
The momentum and pressure equations have the following general form
\begin{equation}
A \psi + {\bf B}\cdot \nabla\psi + C \Delta\psi = f,
\label{elliptic}
\end{equation}
where $A, {\bf B}$ and $C$ are known quantities. 
This equation is solved 
with Dirichlet or Neumann boundary conditions 
\begin{equation}
\psi = \psi_{\Gamma D}  \quad \quad \quad \mbox{or} \quad 
\frac{\partial\psi}{\partial\vec{n}} = \psi_{\Gamma N}. \label{ellipticbc}
\end{equation} 

{\bf Remark}: For the x component of the momentum equations we have $A=1, {\bf B} = -\frac{dt}{\tilde\rho}\nabla\tilde\mu, C=-\frac{dt}{\tilde\rho}\tilde\mu$    and $f$ 
is equal to the right hand side of Eq. \ref{intu}. Similarly, for the pressure equation Eq. \ref{poisson1} we have $A=0, {\bf B}=\frac{\nabla\tilde\rho}{\tilde\rho}, C = 1$ and 
$f=\tilde\rho \frac{\nabla\cdot{\bf v}^{*}}{dt}$. 

In the following section we describe the method of solving equations  Eqs. \ref{elliptic} - \ref{ellipticbc} by a meshfree particle method, called the Finite 
Pointset Method (FPM). 
 
\subsection{ A meshfree particle method for general elliptic boundary value problems}
\label{der_approx}
In this subsection we describe a meshfree method for solving  second order elliptic boundary value problems of type Eqs. \ref{elliptic} - \ref{ellipticbc}. 
The method will be described in a two-dimensional space.  The extension of the method to three-dimensional space is straightforward. 
Let  $\Omega\in R^2$ be the computational domain. The domain $\Omega$ is approximated by particles of positions ${\bf x}_i, i=1,\ldots,N$,  which are socalled numerical grid points. Consider a scaler function $\psi({\bf x})$ and let $\psi_i = \psi({\bf x}_i)$ be its discrete values at particle indices $i=1,\ldots, N$. 
We approximate the spatial derivatives of  $\psi({\bf x})$ at an arbitrary position ${\bf x} \in \{ {\bf x}_i, i = 1, \ldots, N  \}$,  from the values of its neighboring points.   
We introduce a  weight function $w = w({\bf x}_i- {\bf x}, h)$ with a compact support $h$. The value of $h$ can be $2.5$ to  $3$ times the initial spacing of particles such that the minimum number of neighbor is guaranteed in order to approximate the spatial derivatives. But it is user defined quantity. 
This weight function has two properties, first, it avoids the influence of the far particles and the second it reduce the unnecessary neighbors in the computational part.  One can consider different weight function, in this paper 
we consider the Gaussian weight function defined in (\ref{weight}),  where $ \alpha $ is equal to
$6.25$.   Let $ P({\bf x}, h) = \{ {\bf x}_j :j=1,2,\ldots,m \} $ be the
set of $ m $ neighboring particles of $ {\bf x} $ in a circle of radius $h$. We note that the point ${\bf x}$ is itself one of ${\bf x}_j$. 
 
 We consider Taylor expansions of $\psi ({\bf x}_i)$ around $ {\bf x} = (x,y)$
\begin{eqnarray}
\psi(x_j,y_j)= \psi(x,y)+\frac{\partial \psi}{\partial x} (x_j - x) + 
\frac{\partial \psi}{\partial y} (y_j - y) + 
\frac{1}{2} \frac{\partial ^2 \psi}{\partial x^2} (x_j - x)^2 + 
\nonumber 
\\
\quad \quad \quad \frac{\partial ^2 \psi}{\partial x\partial y} (x_j - x)(y_j-y) + 
\frac{1}{2} \frac{\partial^2 \psi}{\partial y^2} (y_j - y)^2 + e_j
\label{taylor}
\end{eqnarray}
for $j = 1, \ldots, m$,
where $ e_j $ is the residual error. 
Let the coefficients of the Taylor expansion be denoted by 
\begin{center}
$
a_1 = \psi(x,y), \;
a_2 = \frac{\partial\psi}{\partial x}, \;
a_3 = \frac{\partial\psi}{\partial y}, \; $\\
$
a_4 = \frac{\partial^2\psi}{\partial x^2}, \;
a_5 = \frac{\partial^2\psi}{\partial x\partial y}, 
a_6 = \frac{\partial^2\psi}{\partial y^2}. \;
$
\end{center}
We add the constraint that at particle position $(x,y)$ the partial differential equation (\ref{elliptic}) 
should be satisfied. If the point $(x,y)$ lies on the boundary, also the boundary condition (\ref{ellipticbc}) needs to be satisfied. Therefore,  we add  
Eqs. \ref{elliptic} and \ref{ellipticbc} to the $m$ equations (\ref{taylor}). Equations \ref{elliptic} and \ref{ellipticbc} are re-expressed as 
\begin{eqnarray}
\label{constraint1}
A a_1 + B_1 a_2 + B_2 a_3 + C (a_4 + a_6 ) = f + e_{m+1}\\
\label{constraint2}
n_x a_2 + n_y a_3  = \psi_{\Gamma N} + e_{m+2},
\end{eqnarray}
where ${\bf B} = (B_1, B_2)$ and  $n_x, n_y$ are the $x,y$ components of the unit
normal vector ${\bf n}$ on the boundary $\Gamma$.
 The coefficients $a_i, i = 1,\ldots,6$ are the unknowns. 

We have six unknowns and $m+1$ equations for the interior points and $m+2$ unknowns for the Neumann boundary points.
This means,  we  always need a minimum of six neighbors.  In general, we have  more than six neighbors, so the  system is 
overdetermined and can be written in matrix form as 
\begin{equation}
{\bf e}= M {\bf a} -  {\bf b},
\label{error}
\end{equation}
where 
%\begin{displaymath}
\begin{eqnarray}
M=
\left( \begin{array}{cccccc}
 1 & ~dx_1 & ~dy_1 & ~\frac{1}{2}dx^2_1 & ~dx_1 dy_1 & ~\frac{1}{2} dy^2_1    \\
\vdots & \vdots  &\vdots & \vdots &\vdots &\vdots   \\
1  &~dx_m & ~dy_m & ~\frac{1}{2}dx^2_m  & ~dx_m dy_m & ~\frac{1}{2} dy^2_m \\
A & ~B_1 & ~B_2 & ~C & ~0 & ~C \\
0 & ~n_x  & ~n_y & ~0  &~0  &~0
 \end{array} \right), 
 \label{matrixM1}
\end{eqnarray}
%\end{displaymath}
with the vectors given by   ${ \bf a} = \left ( a_1, a_2  , \ldots   a_6 \right )^T , \;
{\bf b} =  \left ( \psi_1 , \ldots , \psi_m, f, \psi_N \right )^T $ and
${ \bf e} = \left ( e_1, \ldots, e_m, e_{m+1}, e_{m+2} \right )^T $ 
 and
$dx_j = x_{j} - x, \;  dy_j = y_{j}-y$. 
For the numerical implementation, we set $n_x = n_y = 0$ and  $\psi_{\Gamma N} = 0$ for the interior particles. 
For the Dirichlet boundary particles, we directly prescribe the boundary conditions, and for the 
Neumann boundary particles the matrix coefficients are given by Eq. \ref{matrixM1}.  
The unknowns $a_i$ are computed by minimizing a weighted error over
the neighboring points. Thus, we have to minimize the following quadratic form
\begin{equation}
J = \sum_{i=1}^{m + 2} w_i e_i^2  = (M {\bf a} - {\bf b})^T W (M {\bf a} - {\bf b}), 
\label{functional}
\end{equation}
where 
\begin{eqnarray*}
W=\left( \begin{array}{cccccc}
w_1 & 0 & \cdots& 0 & 0 & 0 \\
\vdots & \vdots & \cdots & \vdots \\
0 & 0 & \cdots & w_m  &0 & 0\\
0 & 0 & \cdots & 0  &1 & 0 \\
0 & 0 & \cdots & 0  &0 & 1
\end{array} \right).
\end{eqnarray*}
The minimization of $ J $ with
respect to ${\bf a}$ formally yields ( if $M^T W M$ is nonsingular)
\begin{equation}
{\bf a} = (M^T W M)^{-1} (M^T W) {\bf b}.
\label{lssol}
\end{equation}
 
In  Eq. \ref{lssol} the vector $( M ^T  W) { \bf b}$ is explicitly  given by
\begin{eqnarray}
( M ^T  W) { \bf b} =
\left( \sum_{j=1}^m w_j \psi_j , \;
\sum_{j=1}^m w_j dx_j \psi_j + B_1 f + n_x \psi_{\Gamma N},
\right.  
\nonumber
\\ \left.
\sum_{j=1}^m w_j dy_j \psi_j + B_2 f + n_y \psi_{\Gamma N}, \;  
\frac{1}{2}\sum_{j=1}^m w_j dx^2_j \psi_j + C f , \;
\right.
\nonumber 
\\ \left.
\sum_{j=1}^m w_j  dx_j dy_j   \psi_j,  \;
\frac{1}{2}\sum_{j=1}^m w_j dy^2_j \psi_j + C f \;
\right )^T.
\end{eqnarray}

Equating the first components on both sides of Eq. \ref{lssol}, we get 
\begin{eqnarray}
\psi = Q_{1} \left(\sum_{j=1}^m w_j \psi_j  \right) +
Q_{2} \left ( \sum_{j=1}^m w_j dx_j \psi_j + B_1 f + n_x \psi_{\Gamma N} \right) +
\nonumber
\\
Q_{3}\left(\sum_{j=1}^m w_j dy_j \psi_j + B_2 f + n_y \psi_{\Gamma N} \right) +
Q_{4} \left( \frac{1}{2}\sum_{j=1}^m w_j dx^2_j \psi_j + C f\right) +
\nonumber 
\\
Q_{5} \left(  \sum_{j=1}^m w_j dx_j dy_j \psi_j \right) +
Q_{6} \left( \frac{1}{2}\sum_{j=1}^m w_j dy^2_j \psi_j + C f\right), 
\end{eqnarray}
where $Q_{1}, Q_{2}, \ldots, Q_{6}$ are the components of the first row of
the matrix $( M^T  W  M)^{-1}$.
Rearranging the terms, we have
\begin{eqnarray}
\nonumber
\psi - \sum_{j=1}^m w_j\left ( Q_{1} + Q_{2} dx_j +
Q_{3} dy_j +  Q_{4} \frac{dx^2_j}{2} +
Q_{5} dx_j ~dy_j + Q_{6} \frac{dy^2_j}{2}   \right) \psi_j =
\\ 
\left ( Q_{2} B_1 + Q_{3} B_2  + Q_{4} C
+ Q_{6} C  \right ) f + 
\left(Q_{2} n_x + Q_{3} n_y  \right ) \psi_{\Gamma N}. \quad \quad \quad
\label{sparsesy}
\end{eqnarray}

 We obtain the following sparse linear 
system of equations for the unknowns $\psi_i, i=1,\ldots, N$
\begin{eqnarray}
\nonumber
\psi_i - \sum_{j=1}^{m(i)} w_{i_j}\left ( Q_{1} + Q_{2}  dx_{i_j} +
Q_{3} dy_{i_j}  + Q_{4} \frac{dx^2_{i_j}}{2} +
Q_{5} dx_{i_j} dy_{i_j} +
Q_{6} \frac{dy^2_{i_j}}{2} \right ) \psi_{i_j} = 
\\ 
\left ( Q_{2} B_1 + Q_{3} B_2  + Q_{4} C
+ Q_{6} C  \right ) f_i + 
\left(Q_{2} n_x + Q_{3} n_y  \right ) \psi_{\Gamma N_i}.  \quad \quad \quad
\label{sparsesy1}
\end{eqnarray}
In  matrix form we have
\begin{equation}
\label{sparsesy2}
L{\bf \Psi} = {\bf R},
\end{equation}
where ${\bf R}$ is the right-hand side vector, ${\bf \Psi}$ is the unknown vector and 
$L$ is the sparse matrix having non-zero entries only for neighboring particles.  

We solve the sparse system (\ref{sparsesy2}) by the Gauss-Seidel method.   
In each time iteration the initial values of $\psi$  for time step 
$n+1$ are taken as the values from previous time step $n$.
While solving the equations for intermediate velocities and the pressure will require more iterations in the first few time steps. After a certain number of time steps, the velocities values and the pressure values at the old 
time step are close to those of new time step, so the number of iterations is dramatically reduced.
 
We stop the iteration process if  
\begin{equation}
\frac{\sum_{i=1}^N |\psi_i^{\tau + 1} - \psi_i ^{(\tau )} | }
{\sum_{i=1}^N |\psi^{(\tau + 1)}_i |} < \epsilon, 
\label{error}
\end{equation}
for $\tau = 0, 1, 2, \ldots $
and $ \epsilon $ is a small positive constant and can be 
defined by the user.  

%{\bf Remark}: The computation of the spatial derivatives has similar procedure, only difference is that we remove the additional constraints in the linear system 
%(\ref{lssol}). We refer \cite{DTKB} for details. 

%%%%%%%%%%%%%%%%%%%%%%%%%%%%%%%%%

\section{Bubble cutting study}

\subsection{Validation: low viscosity}
In a first step, we validate the simulations with the experimental results of the single bubble cutting in reversed osmosis purified water. We apply the same bubble diameter from the experiments ($6.5mm$) and the wire diameter of $3mm$ in the simulation. The viscosity of the fluid (water) is  $\mu_l = 0.001 Pa.s$ and the interfacial tension between water and air is  $\sigma=0.072 N/m$.. The  density of water is $\rho_l=998.2 kg/m^3$ and the density of air is approximated by $\rho_g=1 kg/m^3$ and the dynamic viscosity of air is $\mu_g=2e^{-5}$. .  
For the numerical simulations we consider a two-dimensional geometry of size $36mm\times63mm$. 
The initial bubble position has the center at $x=18mm$ and $y=10mm$ and the wire has the center at $x=19.5mm$ and $y=45mm$ as shown in Fig. \ref{init_bubble_val}. The bubble is approximated by red particles, the liquid is approximated by blue particles. The white circular spot is the position of the wire. We have considered the 
total number of boundary particles, (including 4 walls and the wire) equal to $527$ and the initial number of interior particles equal to $18310$. The constant time 
step $t=5e^{-6}$ is considered. 
Here the horizontal distance between the initial center of bubble and the wire is $d_x = 1.5mm$.  In all four walls and the wire we have considered no-slip boundary conditions. 
Initially, the velocity and pressure are equal to zero. The gravitational force is  ${\bf g} = (0, -9.81) m/s^2$. \\

The comparison between simulation and experiment is depicted in Fig. \ref{validation1} and \ref{validation2}. We extracted a time sequence from the experiments and the corresponding simulations, starting at 0.19 seconds simulated time to 0.27 seconds. The temporal distance between each image is 0.02 s. The rising bubble approaches the wire and starts to deform. There is no direct contact during this phase between the bubble and the wire. Due to the non central approach to the wire, the bubble is cut in a smaller daughter bubble (right) and a larger bubble (left). The larger bubble has three times the diameter of the smaller bubble. The comparison of the cutting process gives a qualitatively good agreement between the experiment and the simulation. Also the shape and size of the mother and daughter bubbles are qualitatively very good agreement. 
 A  detailed comparison of the bubble path from experiment and simulation is shown in Fig. \ref{bubble_path}.  The bubble position during first contact is important, which agrees well between experiment and simulation. Nevertheless, the path of the larger bubble in the simulation shows after the cutting a slightly different behaviour than in the experiment. In the experiment, the larger bubble moves inwards again, while the bubble in the simulation moves horizontally away from the wire, which may arise from a slight horizontally movement of the bubble in the experiment. The cutting of the bubble also depends on the bubble velocity, plotted in Fig. \ref{bubble_velocity}. The bubble accelerates in the simulation and finally reaches the same end velocity that is observed in the experiment. After the splitting into two daughter bubbles, the larger daughter bubble raises faster than the smaller one. The experimental results are governed by higher fluctuations especially for the smaller bubble, which results from very short temporal distances between the images and fluctuations by detecting the bubble interface. Neverthless, the average velocity for the smaller bubble fits with 0.12 m/s quiet well to the simulated result.

\begin{figure}
\centering
 \includegraphics[keepaspectratio=true, width=.8\textwidth]{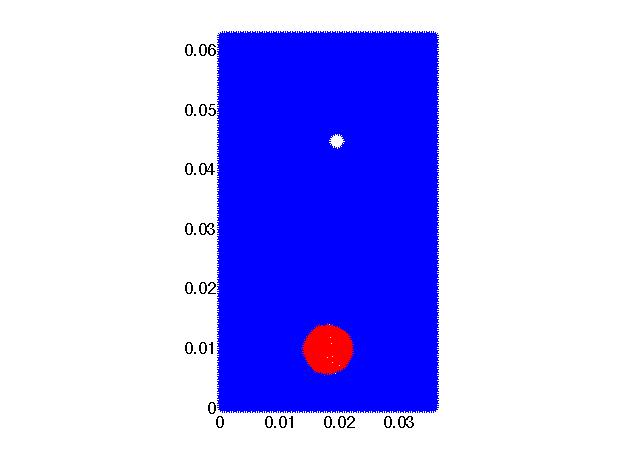}
 \caption{Initial position of bubble and liquid particles for low viscosity.}
 \label{init_bubble_val}
\end{figure}

\begin{figure}
	\captionsetup[subfigure]{margin=5pt} %parskip=0pt, hangindent=0pt, indention=0pt, singlelinecheck=true}
\captionsetup[subfigure]{labelformat=empty}
	\subfloat[$ Experiment: t = 0.19 s $]{
		\includegraphics[keepaspectratio=true, width=.48\textwidth]{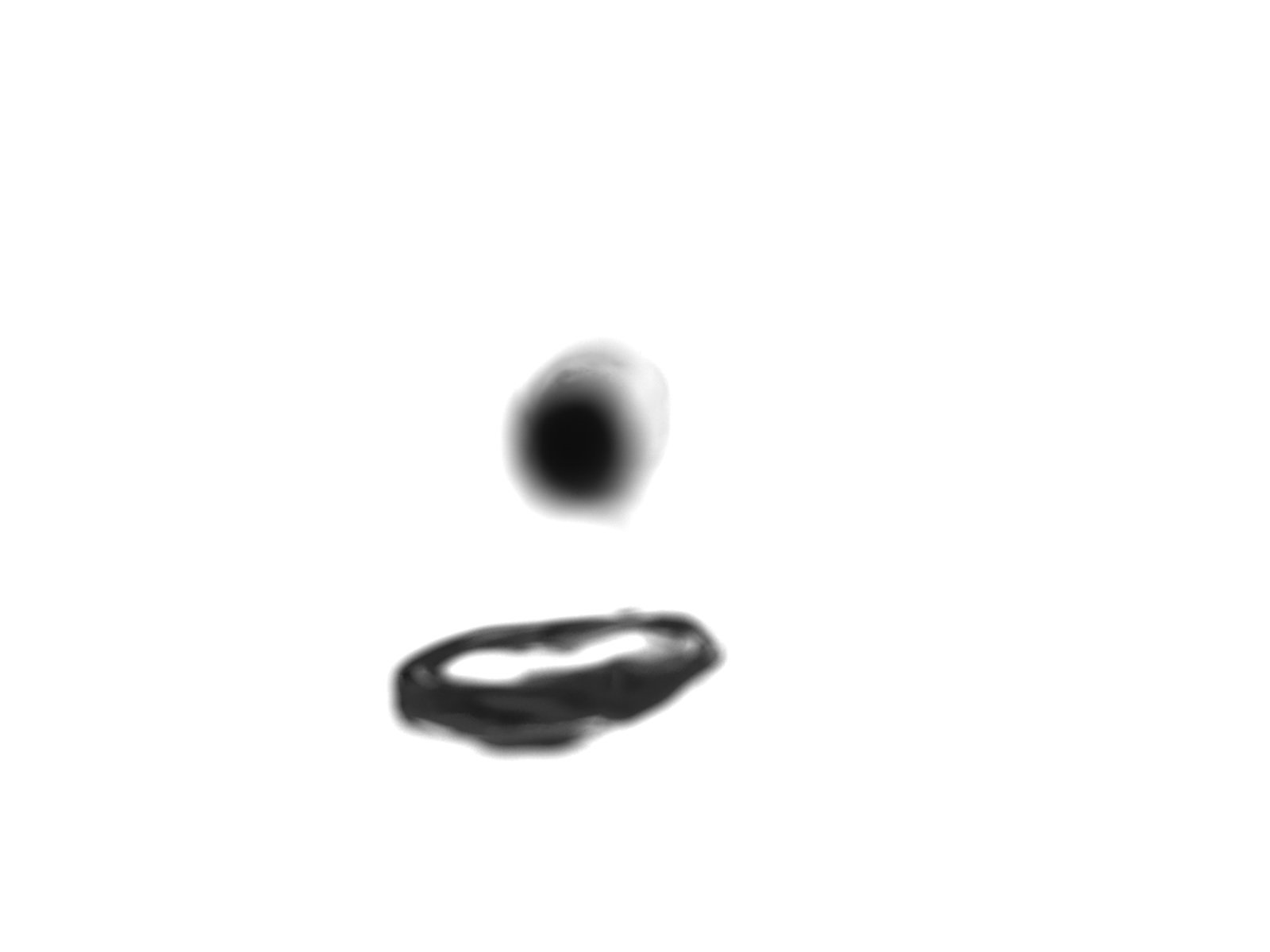}
 	} \subfloat[ $Simulation: t = 0.19 s$]{
		\includegraphics[keepaspectratio=true, width=.48\textwidth]{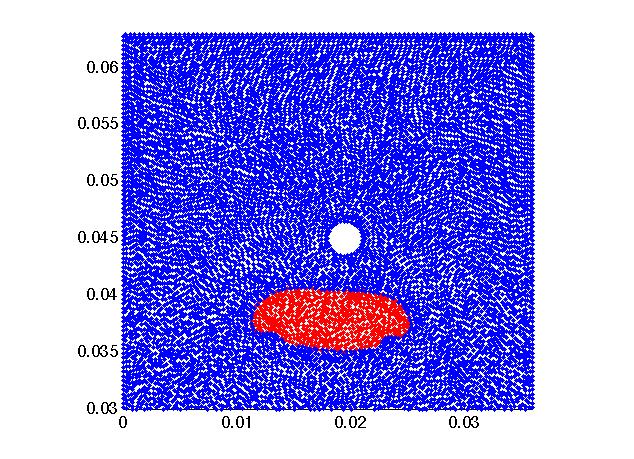}
	} \\
	\captionsetup[subfigure]{margin=0pt} %parskip=0pt, hangindent=0pt, indention=0pt, singlelinecheck=true}
	\subfloat[$ Experiment: t = 0.21 s $]{	
		\includegraphics[keepaspectratio=true, width=.48\textwidth]{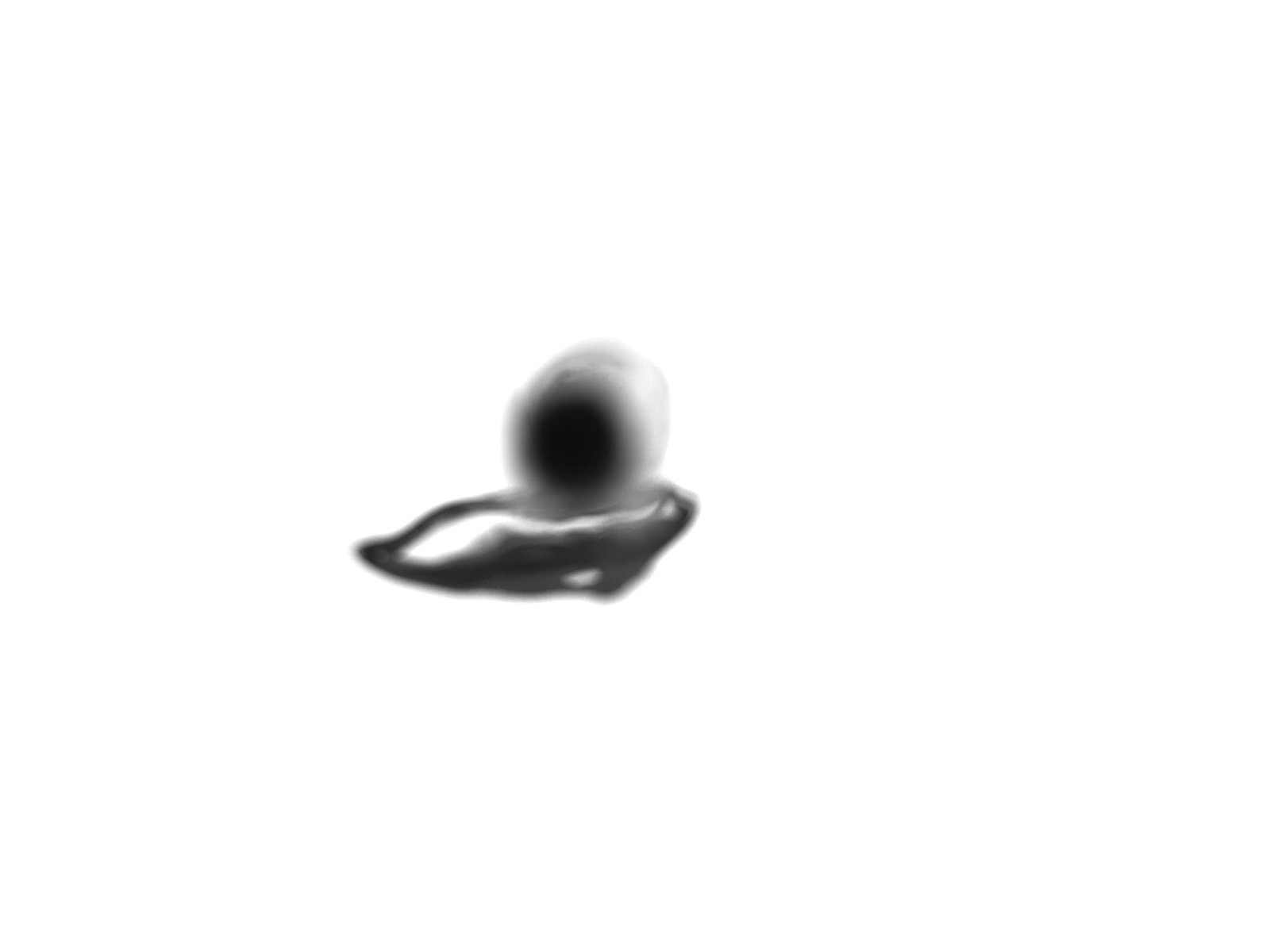}
 	} \subfloat[ $Simulation: t = 0.21 s$]{
		\includegraphics[keepaspectratio=true, width=.48\textwidth]{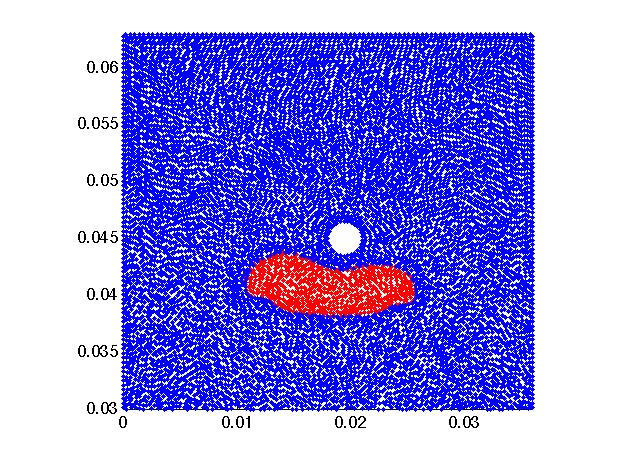}
 	}   \\
	\captionsetup[subfigure]{margin=0pt} %parskip=0pt, hangindent=0pt, indention=0pt, singlelinecheck=true}
	\subfloat[$ Experiment: t = 0.23 s $]{
		\includegraphics[keepaspectratio=true, width=.48\textwidth]{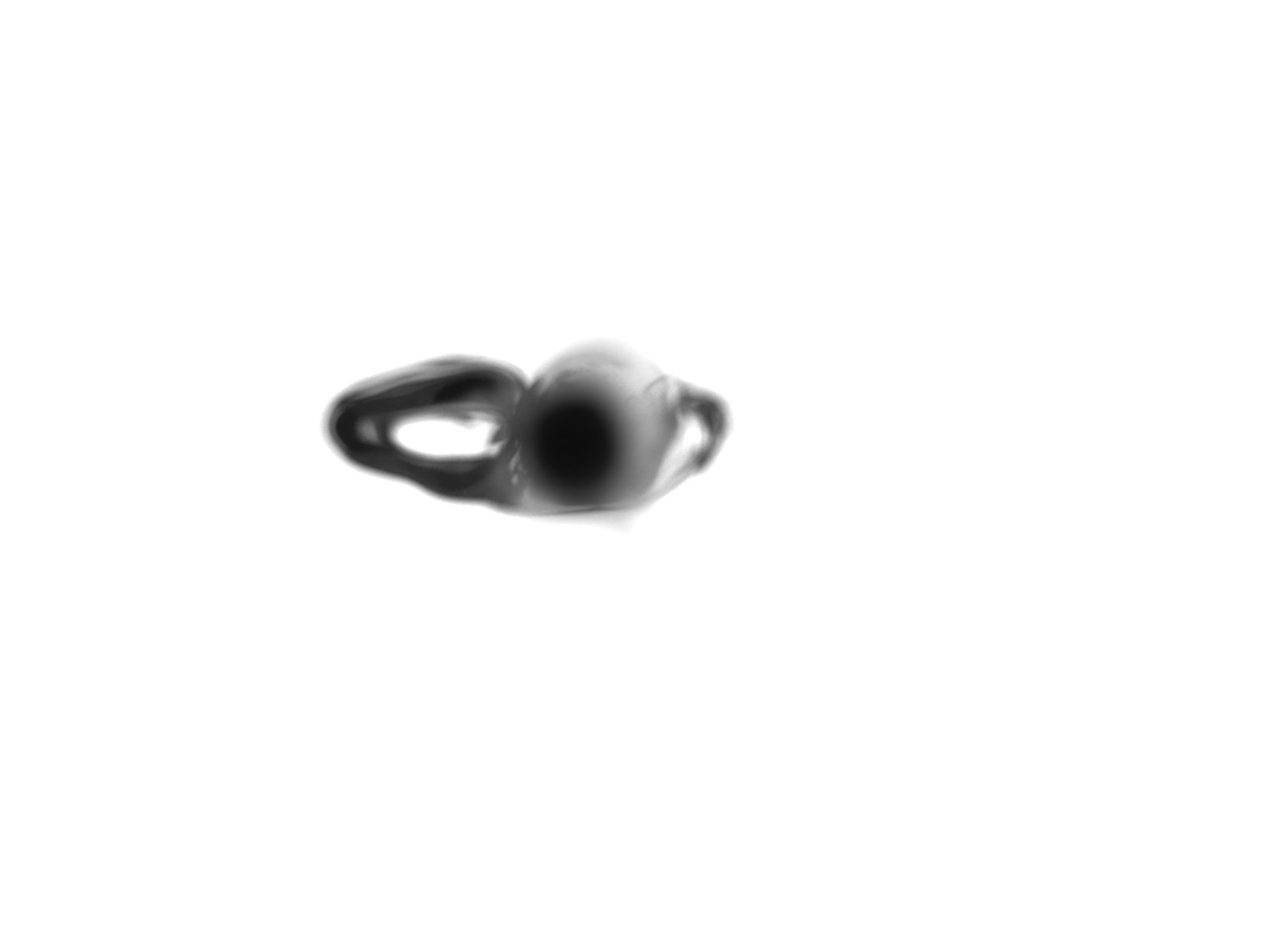}
 	} \subfloat[ $Simulation: t = 0.23 s$]{
		\includegraphics[keepaspectratio=true, width=.48\textwidth]{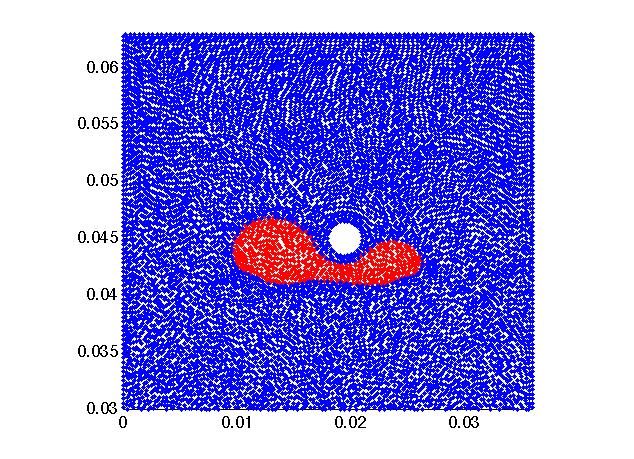}
 	}  
 	
 		\captionsetup{margin=20pt}
 		\caption{Bubble cutting over time $t=0.19s, 0.21s, 0.23s$. Experiment (left) vs. simulation (right).}
 		\label{validation1}
 	\end{figure}

 \begin{figure}	
	\captionsetup[subfigure]{margin=0pt} %parskip=0pt, hangindent=0pt, indention=0pt, singlelinecheck=true}
	\subfloat[$ Experiment: t = 0.25 s $]{
		\includegraphics[keepaspectratio=true, width=.48\textwidth]{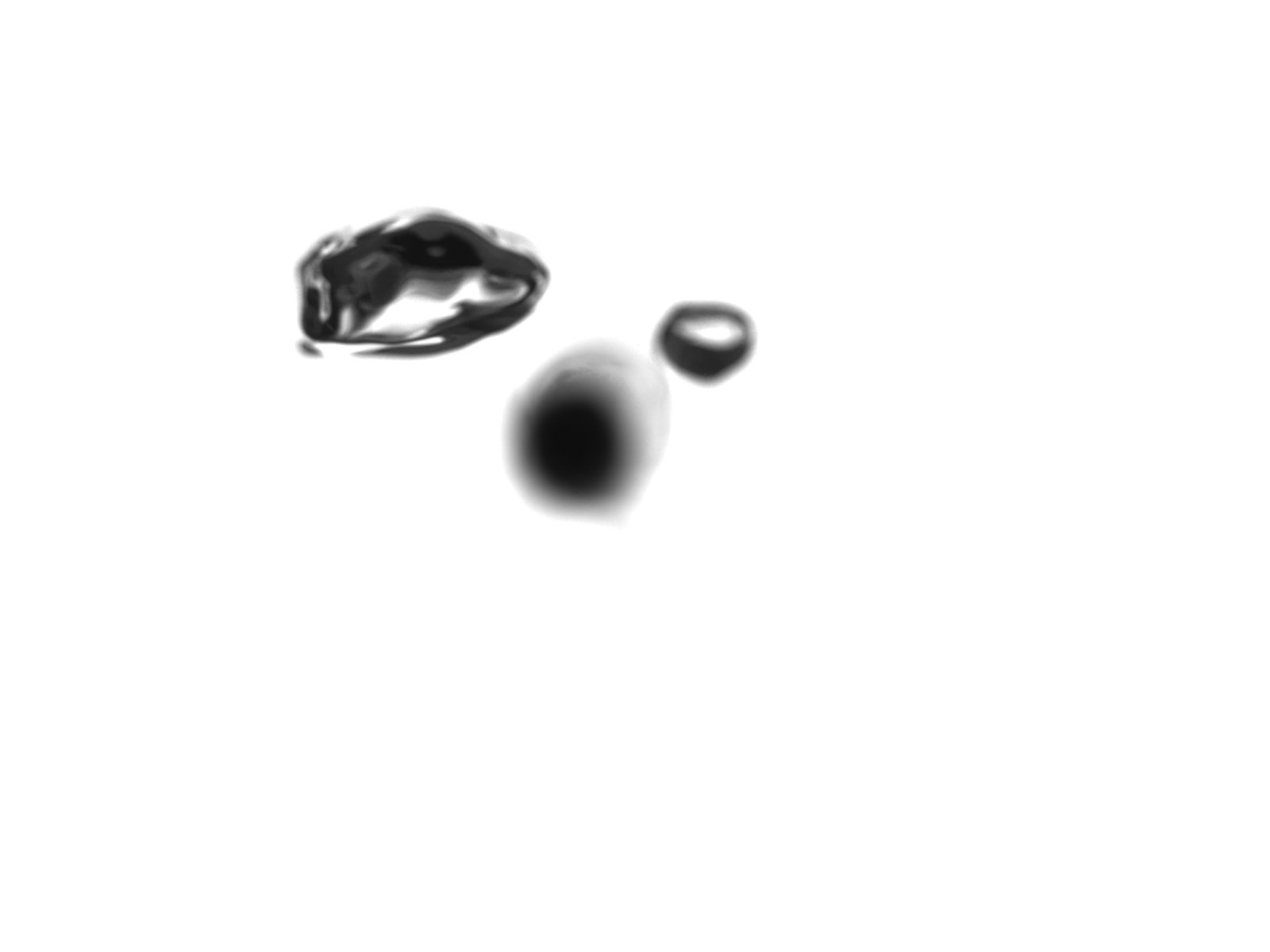}
 	} \subfloat[ $Simulation: t = 0.25 s$]{
		\includegraphics[keepaspectratio=true, width=.48\textwidth]{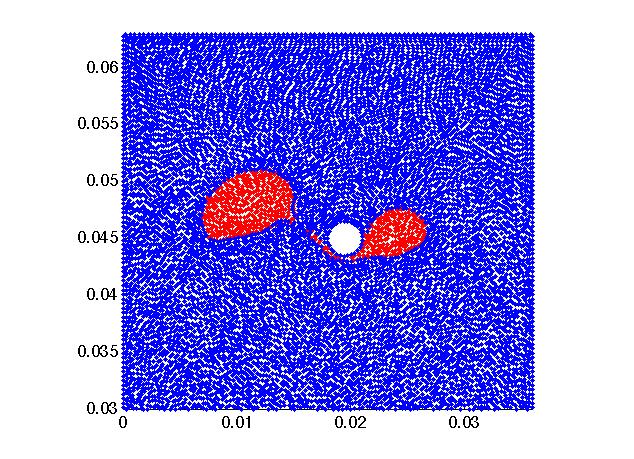}
  }\\
	\captionsetup[subfigure]{margin=0pt} %parskip=0pt, hangindent=0pt, indention=0pt, singlelinecheck=true}
	\subfloat[$ Experiment: t = 027 s $]{
		\includegraphics[keepaspectratio=true, width=.48\textwidth]{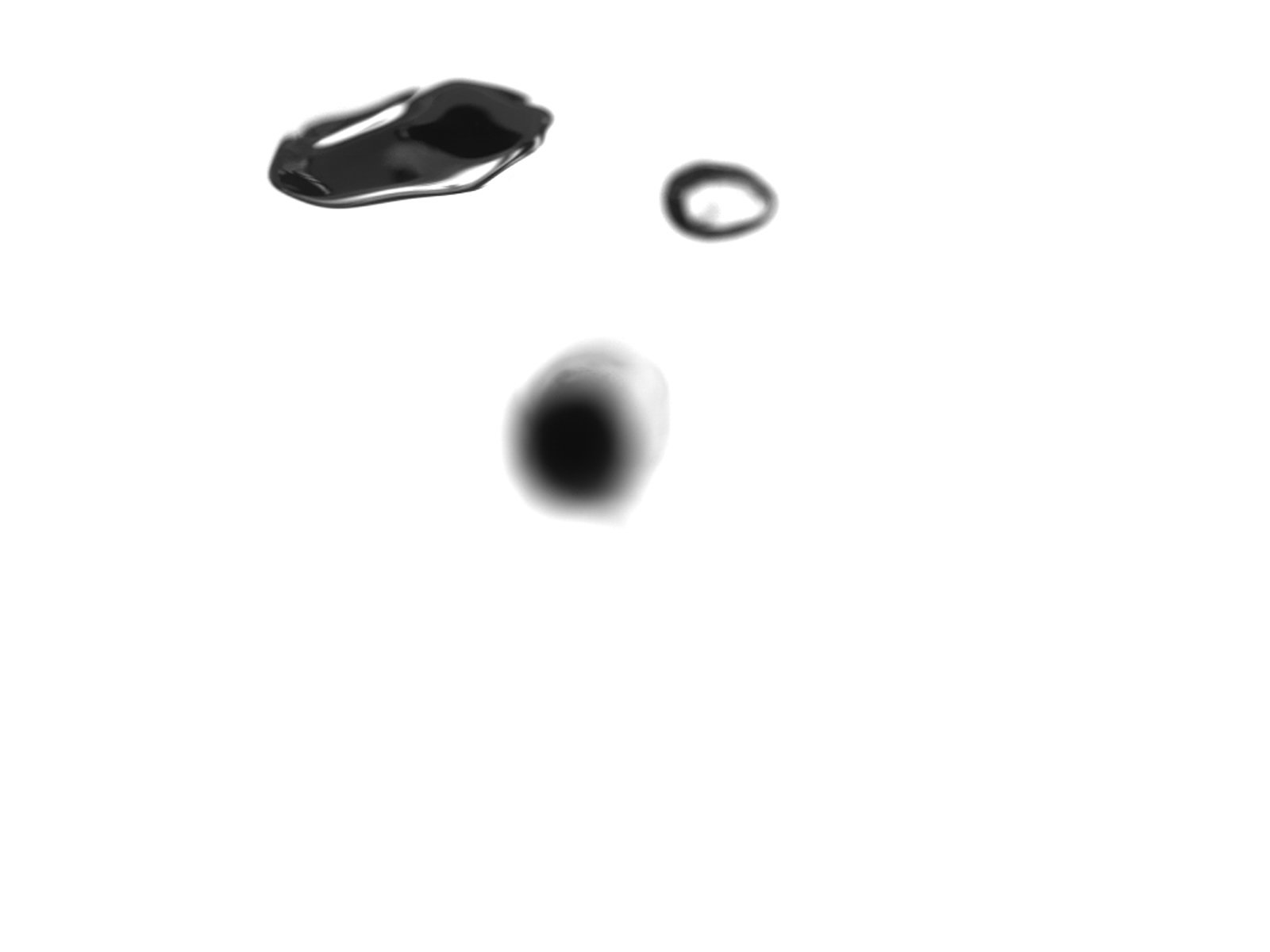}
 	} \subfloat[ $Simulation: t = 0.27 s$]{
		\includegraphics[keepaspectratio=true, width=.48\textwidth]{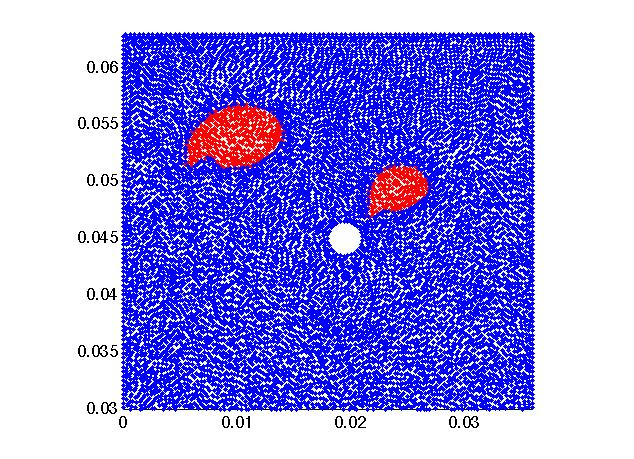}
 	
 	}

	\captionsetup{margin=20pt}
	\caption{Bubble cutting over time $t=0.25s, 0.27s$.. Experiment (left) vs. simulation (right).}
	\label{validation2}
\end{figure}

\begin{figure}
	\captionsetup[subfigure]{margin=0pt} %parskip=0pt, hangindent=0pt, indention=0pt, singlelinecheck=true}
	\subfloat[$ Experiment $]{ \includegraphics[keepaspectratio=true, width=.48\textwidth]{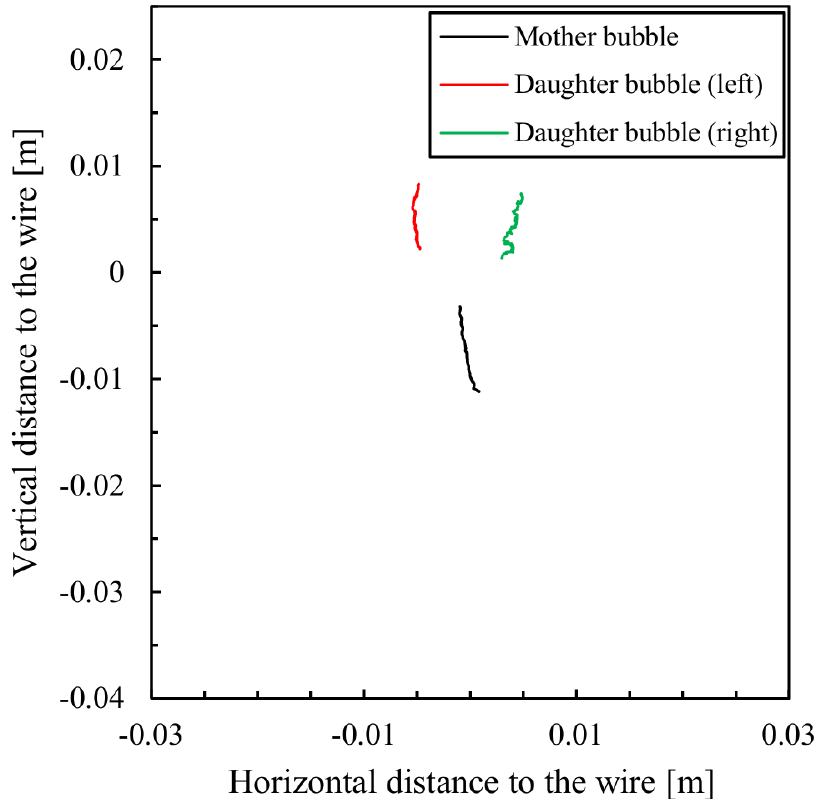}
 	} \subfloat[ $Simulation$]{	
  \includegraphics[keepaspectratio=false, width=.48\textwidth, height=180px]{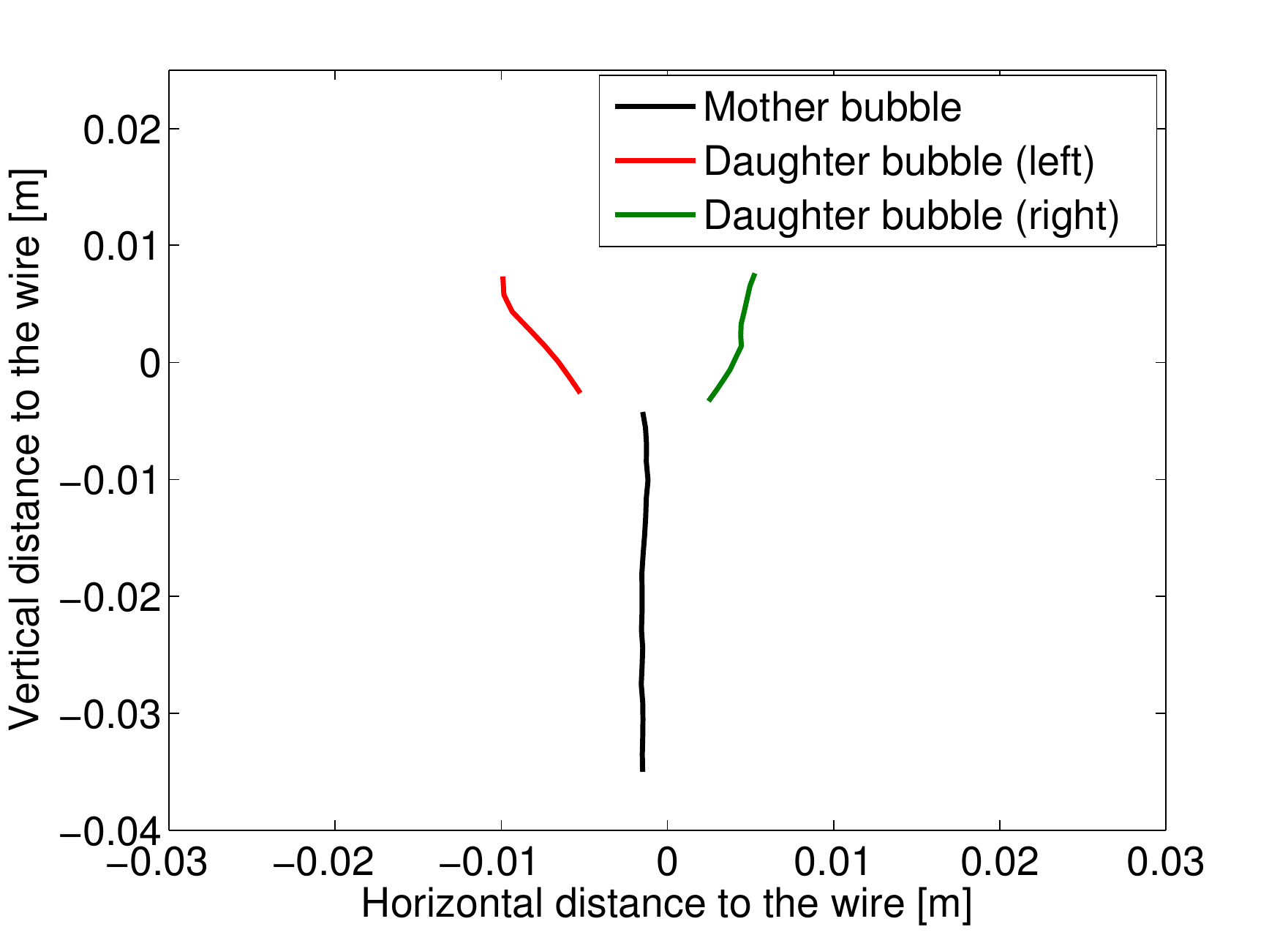}
}
 \caption{Path of the mother bubble and the two daughter bubbles.}
 \label{bubble_path}
\end{figure}

%\begin{figure}
%\centering
% \includegraphics[keepaspectratio=true, width=.48\textwidth]{figures/Experiment/bubble_path.pdf}
%  \includegraphics[keepaspectratio=true, width=.48\textwidth]{figures/Benchmark_bubble_paths_simul.eps}
% \caption{Path of the mother bubble and the two daughter bubbles.}
% \label{bubble_path}
%\end{figure}
%
%\begin{figure}
%\centering
% \includegraphics[keepaspectratio=true, width=.8\textwidth]{figures/Experiment/bubble_velocity.pdf}
% \caption{Velocity of the bubble.}
% \label{bubble_velocity}
%\end{figure}

\begin{figure}
	\captionsetup[subfigure]{margin=0pt} %parskip=0pt, hangindent=0pt, indention=0pt, singlelinecheck=true}
	\subfloat[$ Experiment $]{ \includegraphics[keepaspectratio=true, width=.44\textwidth]{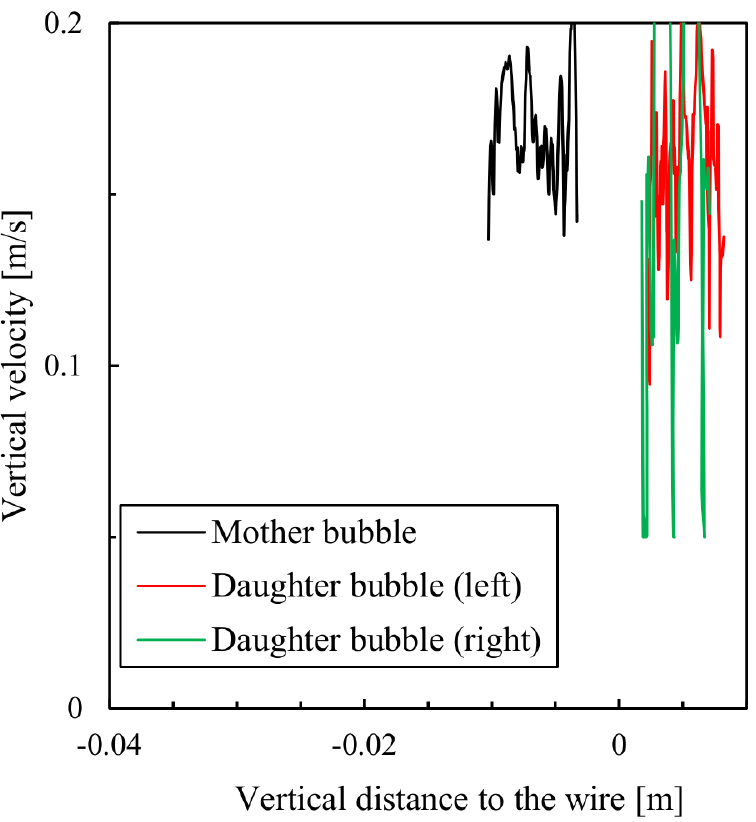}
 	} \subfloat[ $Simulation$]{	
  \includegraphics[keepaspectratio=false, width=.48\textwidth, height=185px]{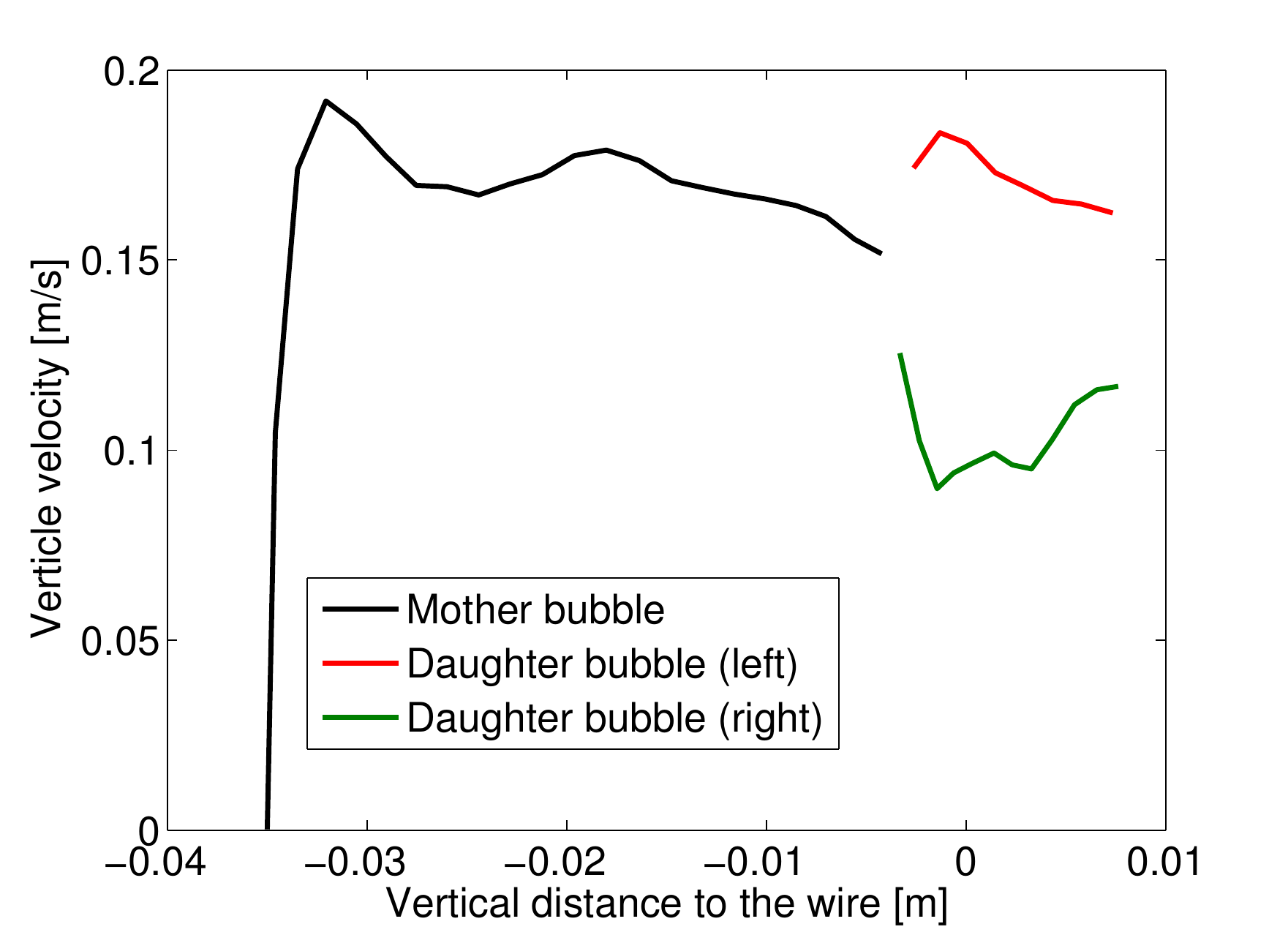}
}
 \caption{Velocities of mother and daughter bubbles.}
 \label{bubble_velocity}
\end{figure}

\subsection{Case study: high viscosity}

The computational domain is the same as in the previous section. However, the position of the wire is changed. 
The data has been taken from chapter 6 of  \cite{Baltussen}. 
The liquid has density $\rho_l=1250 kg/m^3$, dynamical viscosity $\mu_l = 0.219 Pa.s$. Similarly the gas density $\rho_g=1 kg/m^3$ and the viscosity 
$\mu_g=2e^{-5} Pa.s$. The surface tension coefficient $\sigma=0.0658 N/m$.  
 We consider a bubble of diameter $9.14mm$ with its initial center at $(18mm, 9mm)$. 
We consider a  wire (in 2D a circle) of diameter $3.1mm$ with different centers at $y=45mm$ and  $ x=18mm, 18.5mm, 19mm$ and $19.5mm$. 
This means, we consider the initial distance $d_x$  between the center of the bubble and the center of the wire equal to $d_x=0mm, 0.5mm, 1mm, 1.5mm$.  
Fig. \ref{init_bubble} shows the initial  geometry with $d_x = 0mm$. The initial number of particles and the  time step are the same as in the low viscosity case. 
The initial and boundary conditions and the rest of other parameters are same as in the previous case. 
\begin{figure}
\centering
 \includegraphics[keepaspectratio=true, width=.8\textwidth]{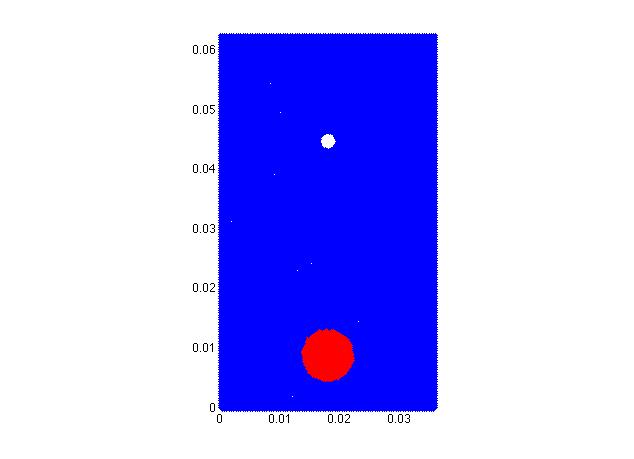}
 \caption{Initial position of bubble and liquid particles with $d_x = 0$.}
 \label{init_bubble}
\end{figure}
 
In Figs. \ref{cut_t_0dot288} - \ref{cut_t_0dot384} we have plotted the positions of the bubble and the wire for $d_x=0mm, 0.5mm, 1mm$ and $1.5mm$ at time $t = 0.288, 0.320, 0.352$ and $t=0.384$ seconds, respectively.  For $d_x = 0$ we observe the wire located in the middle of the bubble as expected. When we increased the distance $d_x$ from $0.5mm$ to $1.5mm$, we observed that the left part of the bubble is increasing and the right part becomes smaller. 
We clearly observe that the daughter bubbles are symmetric for $d_x = 0mm$ in contrast to the other cases.
 We further observe  a small layer between the wire and the bubble. 
After $t=0.352$ seconds we observe the  cutting of the bubble, see Figs. \ref{cut_t_0dot352} and \ref{cut_t_0dot384}. Two daughter bubbles arise, a larger one on the left and a smaller one on the right side of the wire. The overall numerical results are comparable with the results presented in 
\cite{Baltussen}. 
 
%
 
%%%%%%%%%%%%%%%%%%%%%
%%%%%%%%%%%%%%%%%%%%
\begin{figure}
	\captionsetup[subfigure]{margin=5pt} %parskip=0pt, hangindent=0pt, indention=0pt, singlelinecheck=true}
	\subfloat[$ d_x = 0 $]{
		\includegraphics[keepaspectratio=true, width=.48\textwidth]{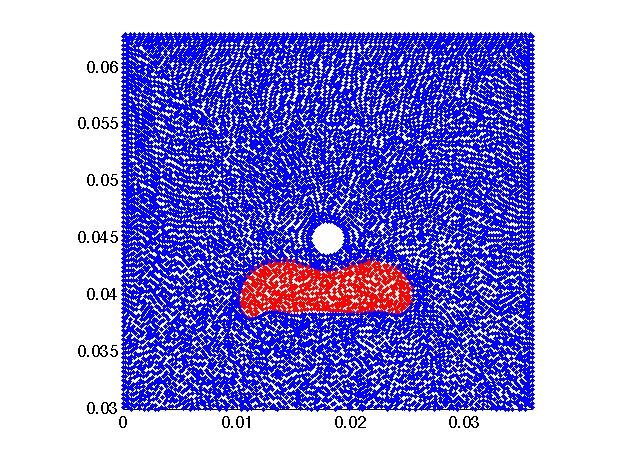}
 	} \subfloat[ $d_x = 0,5$]{
		\includegraphics[keepaspectratio=true, width=.48\textwidth]{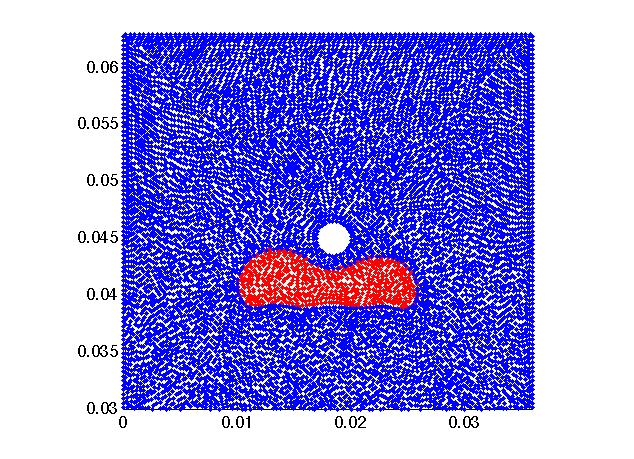}
	} 
	\captionsetup[subfigure]{margin=0pt} %parskip=0pt, hangindent=0pt, indention=0pt, singlelinecheck=true}
	\subfloat[$ d_x = 1 $]{
		\includegraphics[keepaspectratio=true, width=.48\textwidth]{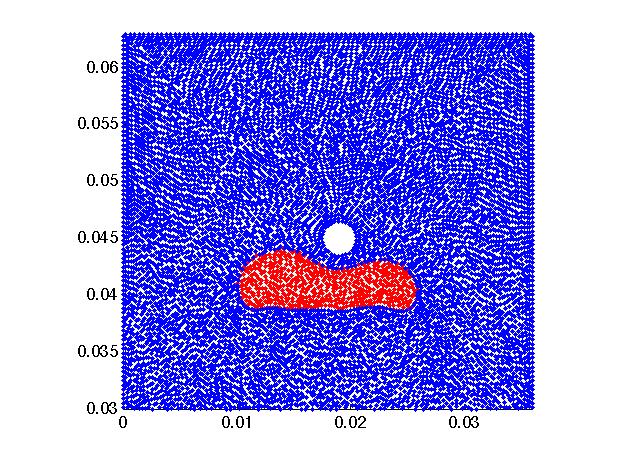}
 	} \subfloat[ $d_x = 1.5$]{
		\includegraphics[keepaspectratio=true, width=.48\textwidth]{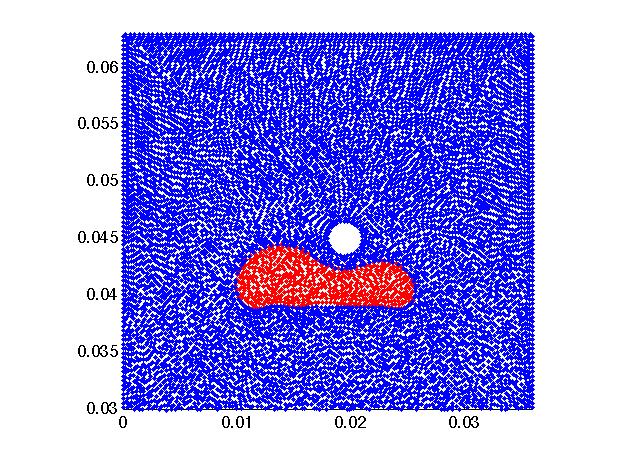} 
 	}   
 	\\
	\captionsetup{margin=20pt}
	\caption{The position of bubble with different positions of the wire at $t=0.288$.  }
	\label{cut_t_0dot288}
\end{figure}

%%%%%%%%%%%%%%%%%%%%
\begin{figure}
	\captionsetup[subfigure]{margin=5pt} %parskip=0pt, hangindent=0pt, indention=0pt, singlelinecheck=true}
	\subfloat[$ d_x = 0 $]{
		\includegraphics[keepaspectratio=true, width=.48\textwidth]{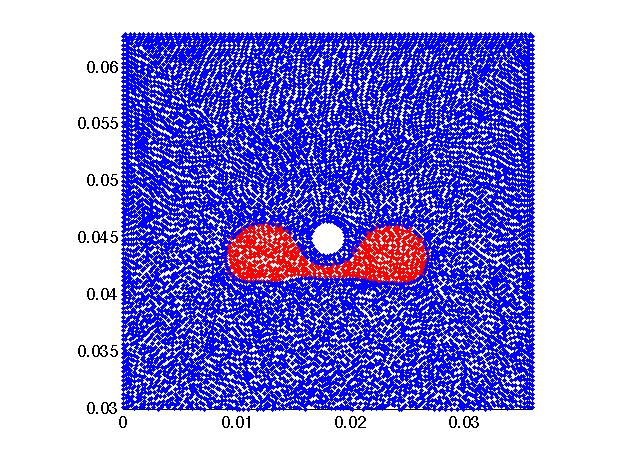}
 	} \subfloat[ $d_x = 0,5$]{
		\includegraphics[keepaspectratio=true, width=.48\textwidth]{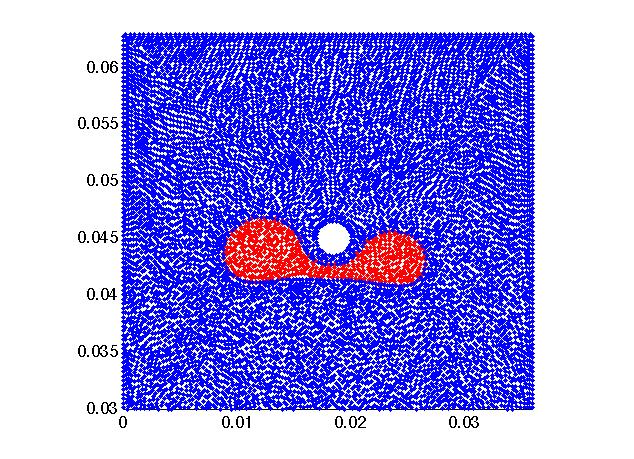}
	} 
	\captionsetup[subfigure]{margin=0pt} %parskip=0pt, hangindent=0pt, indention=0pt, singlelinecheck=true}
	\subfloat[$ d_x = 1 $]{
		\includegraphics[keepaspectratio=true, width=.48\textwidth]{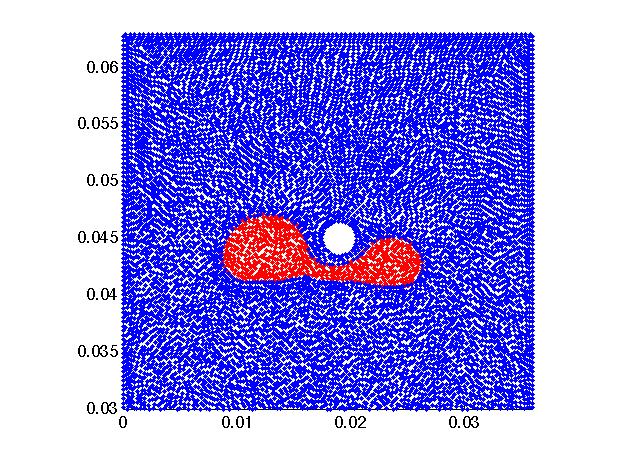}
 	} \subfloat[ $d_x = 1.5$]{
		\includegraphics[keepaspectratio=true, width=.48\textwidth]{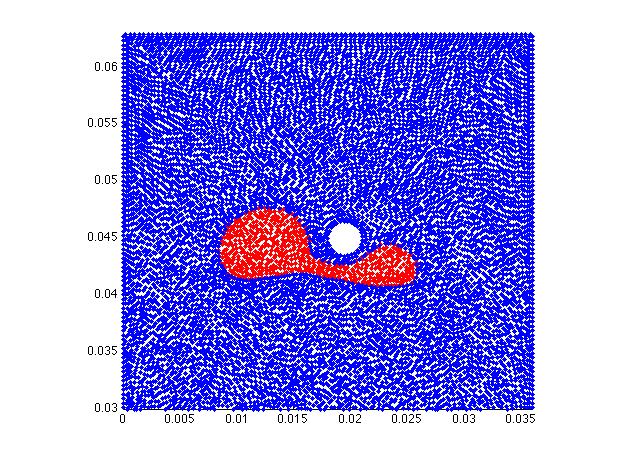} 
 	}   
 	\\
	\captionsetup{margin=20pt}
	\caption{The position of bubble with different positions of the wire at $t=0.320$.  }
	\label{cut_t_0dot320}
\end{figure}

%%%%%%%%%%%%%%%%%%%%%%%%%
%%%%%%%%%%%%%%%%%%%%
\begin{figure}
	\captionsetup[subfigure]{margin=5pt} %parskip=0pt, hangindent=0pt, indention=0pt, singlelinecheck=true}
	\subfloat[$ d_x = 0 $]{
		\includegraphics[keepaspectratio=true, width=.48\textwidth]{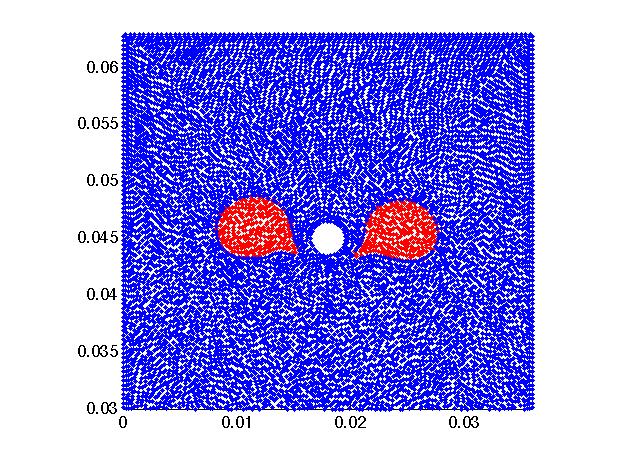}
 	} \subfloat[ $d_x = 0,5$]{
		\includegraphics[keepaspectratio=true, width=.48\textwidth]{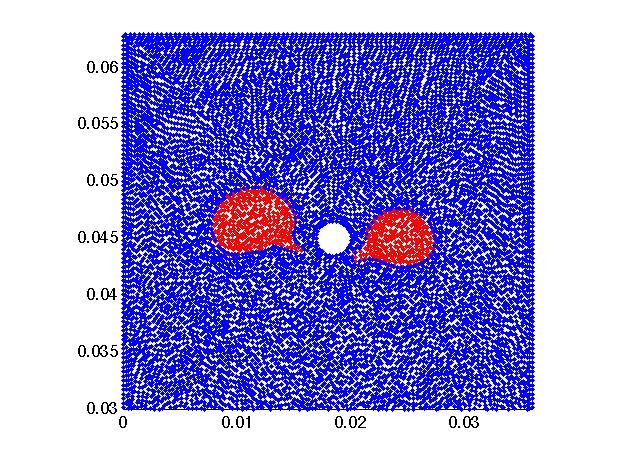}
	} 
	\captionsetup[subfigure]{margin=0pt} %parskip=0pt, hangindent=0pt, indention=0pt, singlelinecheck=true}
	\subfloat[$ d_x = 1 $]{
		\includegraphics[keepaspectratio=true, width=.48\textwidth]{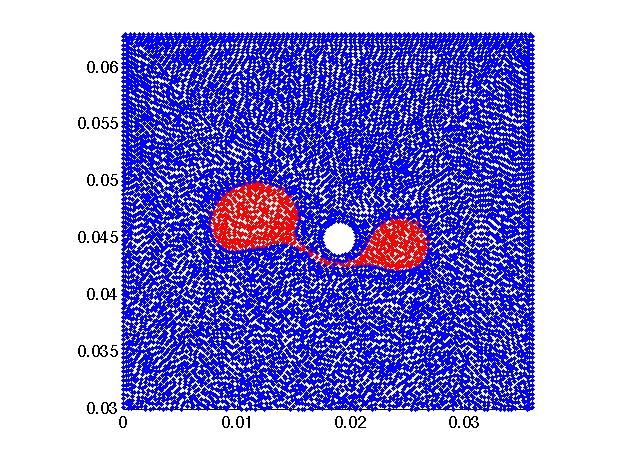}
 	} \subfloat[ $d_x = 1.5$]{
		\includegraphics[keepaspectratio=true, width=.48\textwidth]{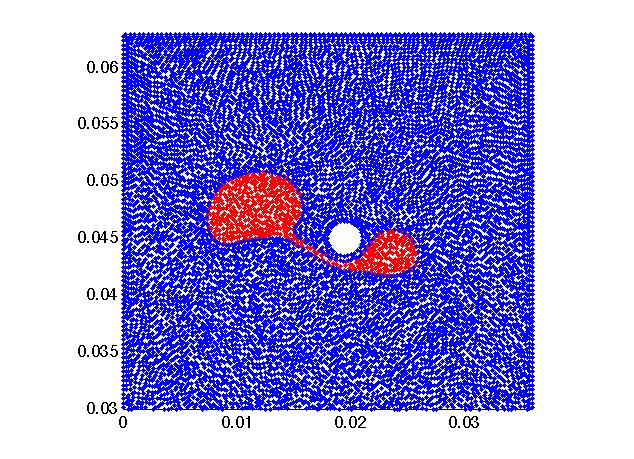} 
 	}   
 	\\
	\captionsetup{margin=20pt}
	\caption{The position of bubble with different positions of the wire at $t=0.352$.  }
	\label{cut_t_0dot352}
\end{figure}

%%%%%%%%%%%%%%%%%%%%
\begin{figure}
	\captionsetup[subfigure]{margin=5pt} %parskip=0pt, hangindent=0pt, indention=0pt, singlelinecheck=true}
	\subfloat[$ d_x = 0 $]{
		\includegraphics[keepaspectratio=true, width=.48\textwidth]{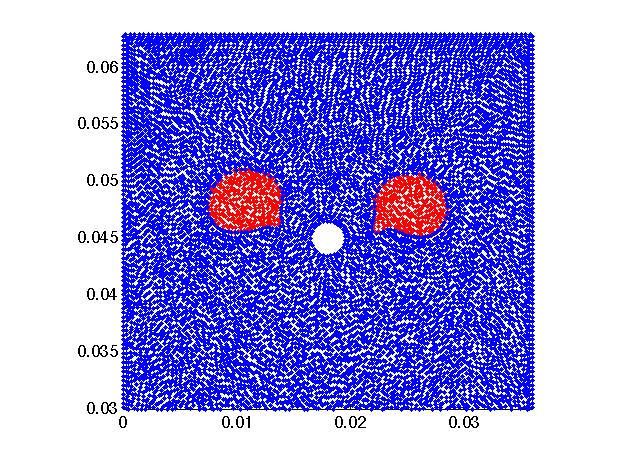}
 	} \subfloat[ $d_x = 0,5$]{
		\includegraphics[keepaspectratio=true, width=.48\textwidth]{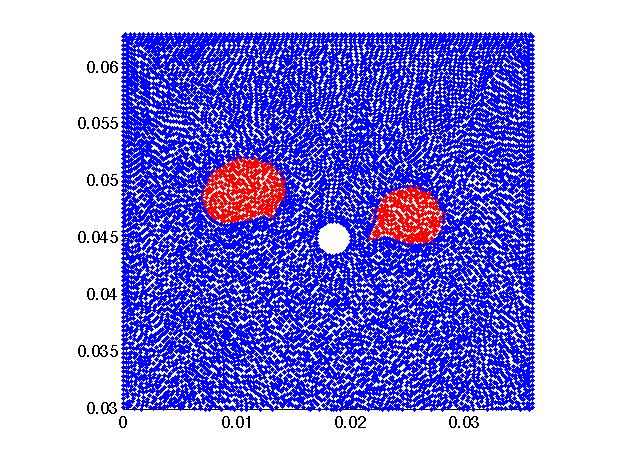}
	} 
	\captionsetup[subfigure]{margin=0pt} %parskip=0pt, hangindent=0pt, indention=0pt, singlelinecheck=true}
	\subfloat[$ d_x = 1 $]{
		\includegraphics[keepaspectratio=true, width=.48\textwidth]{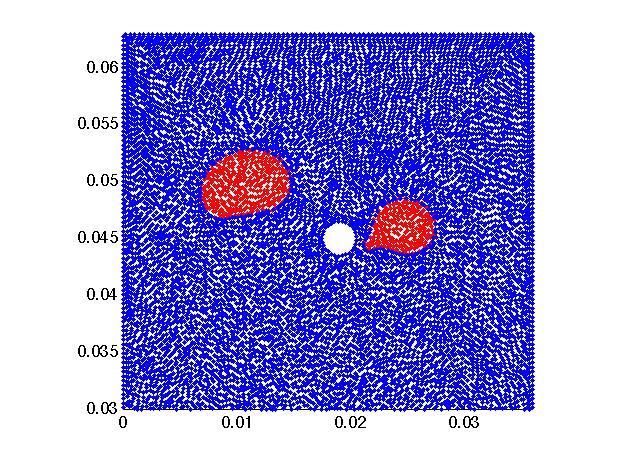}
 	} \subfloat[ $d_x = 1.5$]{
		\includegraphics[keepaspectratio=true, width=.48\textwidth]{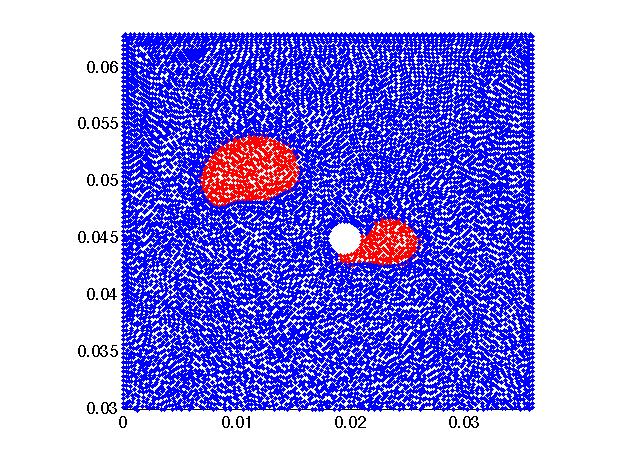} 
 	}   
 	\\
	\captionsetup{margin=20pt}
	\caption{The position of bubble with different positions of the wire at $t=0.384$.  }
	\label{cut_t_0dot384}
\end{figure}

%%%%%%%%%%%%%%%%%%%%
In Fig. \ref{cut_pos} we have plotted the trajectories of the mother and bubble droplets. We observed that the mother droplet is cutted into two daughter 
bubble slightly below the wire, compare with Fig. \ref{cut_t_0dot352}. The trajectories are plotted up to time $t=0.5 s$. We see that when the size of the daughter bubble 
is increasing, it travels longer than the smaller bubbles. The reason is that the rising velocity of the larger bubble is larger than the smaller ones, see 
Fig. \ref{cut_vel} for the velocities of mother and daughter bubbles.

%%%%%%%%%%%%%%%%%%%%%%%%%%%%%%%%%
\begin{figure}
	\captionsetup[subfigure]{margin=5pt} %parskip=0pt, hangindent=0pt, indention=0pt, singlelinecheck=true}
	\subfloat[$ d_x = 0 $]{
		\includegraphics[keepaspectratio=true, width=.48\textwidth]{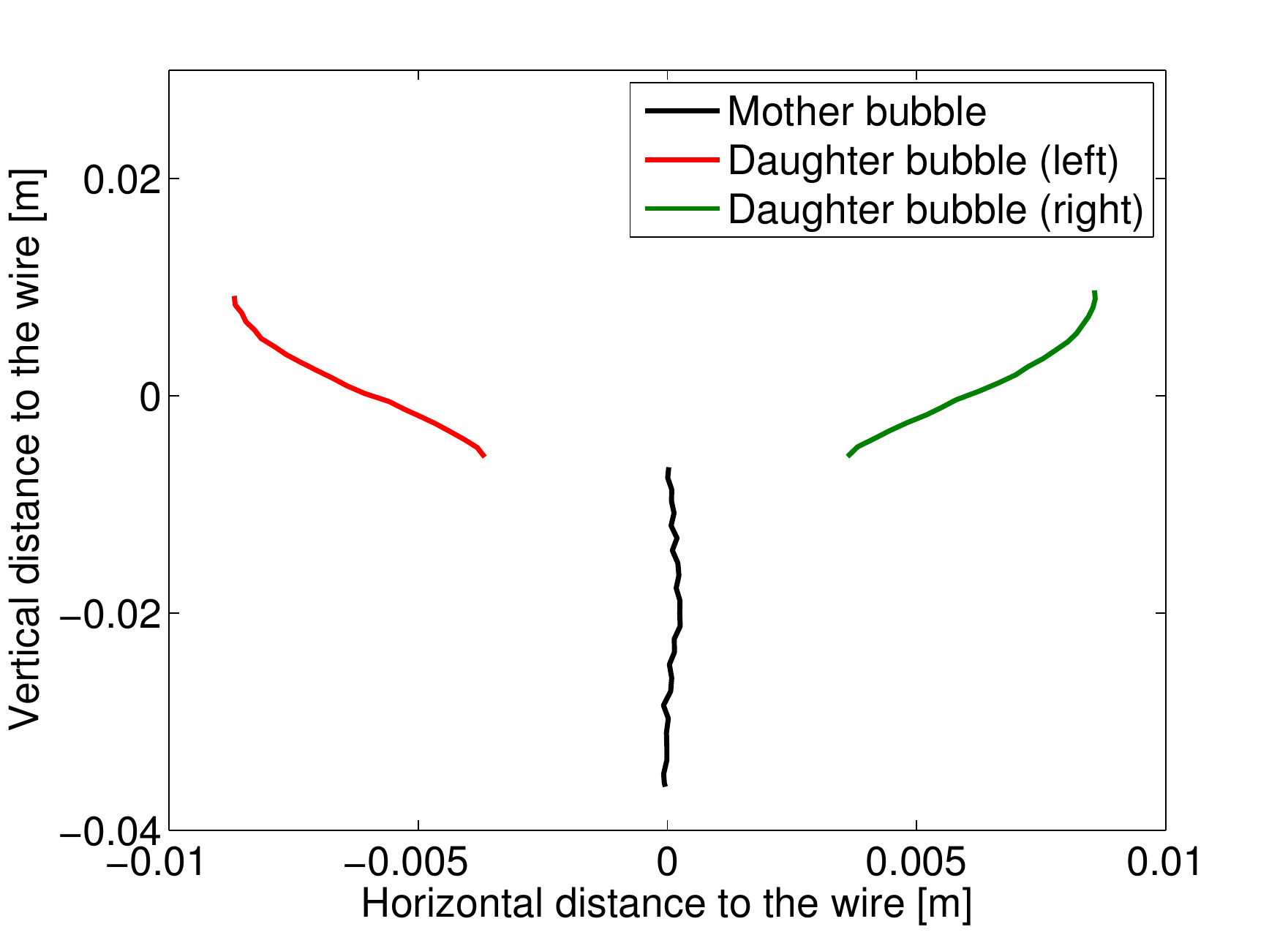}
 	} \subfloat[ $d_x = 0.5$]{
		\includegraphics[keepaspectratio=true, width=.48\textwidth]{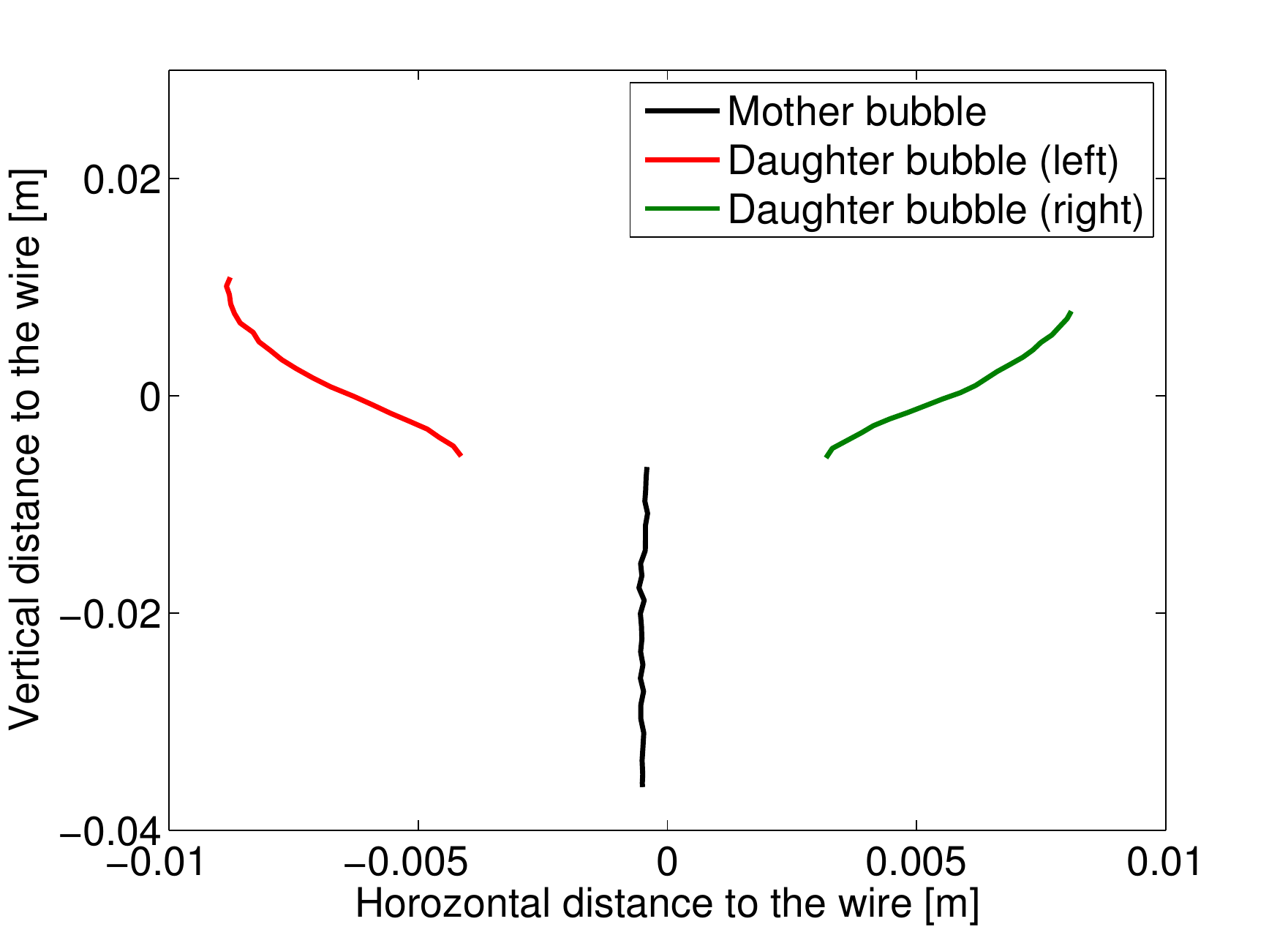}
	} 
	\captionsetup[subfigure]{margin=0pt} %parskip=0pt, hangindent=0pt, indention=0pt, singlelinecheck=true}
	\subfloat[$ d_x = 1 $]{
		\includegraphics[keepaspectratio=true, width=.48\textwidth]{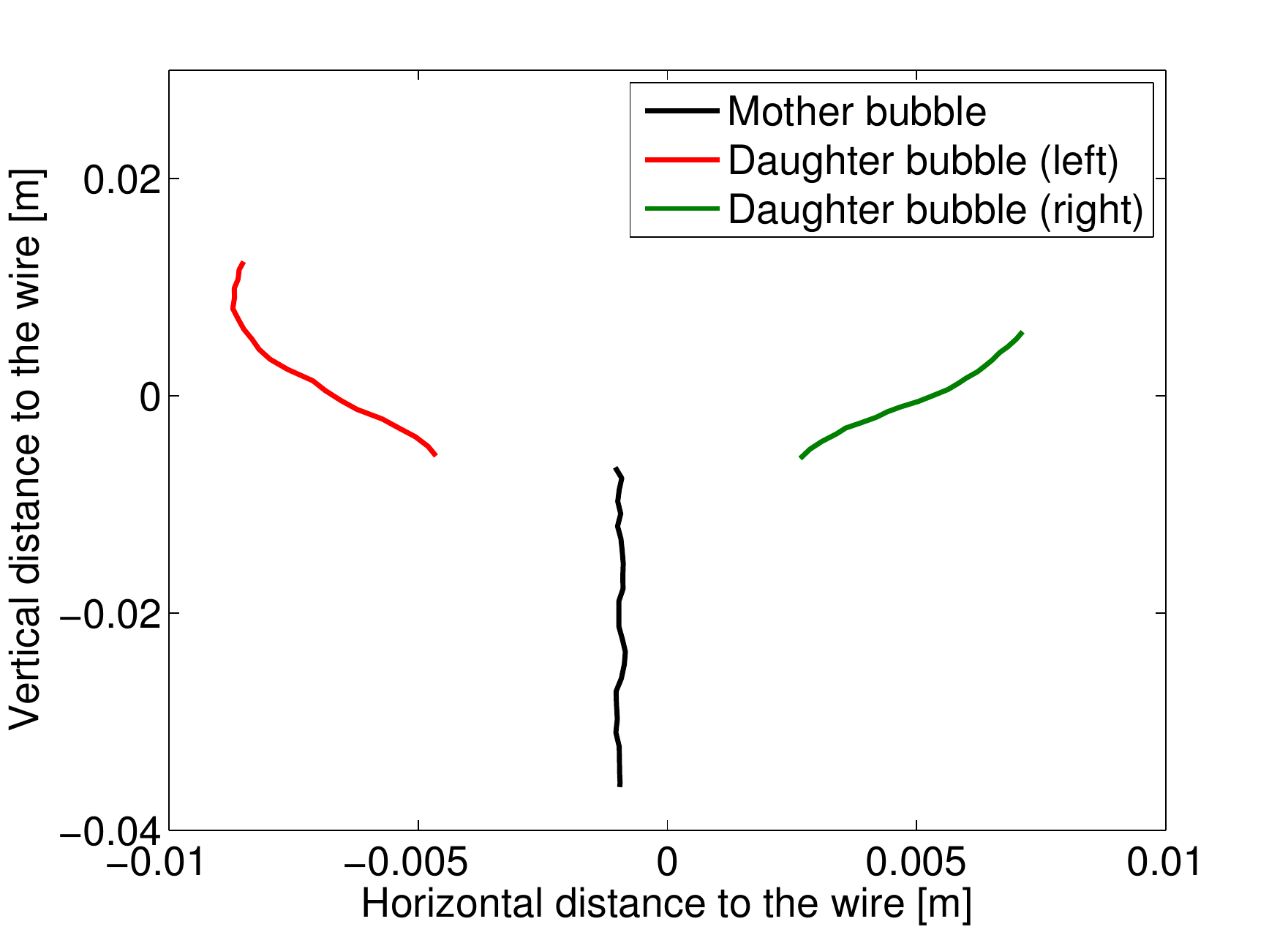}
 	} \subfloat[ $d_x = 1.5$]{
		\includegraphics[keepaspectratio=true, width=.48\textwidth]{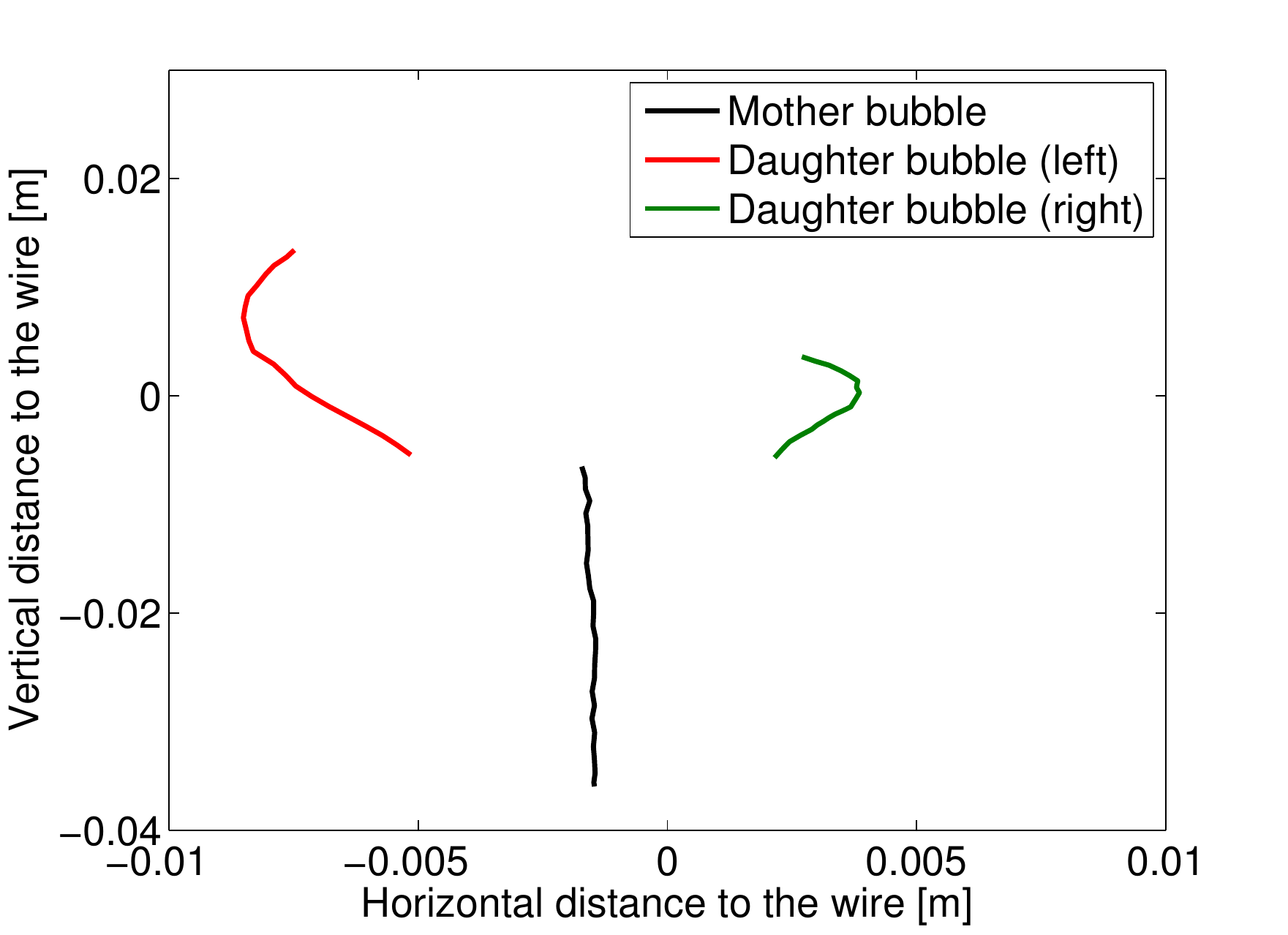} 
 	}   
 	\\
	\captionsetup{margin=20pt}
	\caption{ The trajectories of the mother and the daughter bubbles. . }
	\label{cut_pos}
\end{figure}

%%%%%%%%%%%%%%%%%%%%%%%%%%%%%%%%%%
\begin{figure}
	\captionsetup[subfigure]{margin=5pt} %parskip=0pt, hangindent=0pt, indention=0pt, singlelinecheck=true}
	\subfloat[$ d_x = 0 $]{
		\includegraphics[keepaspectratio=true, width=.48\textwidth]{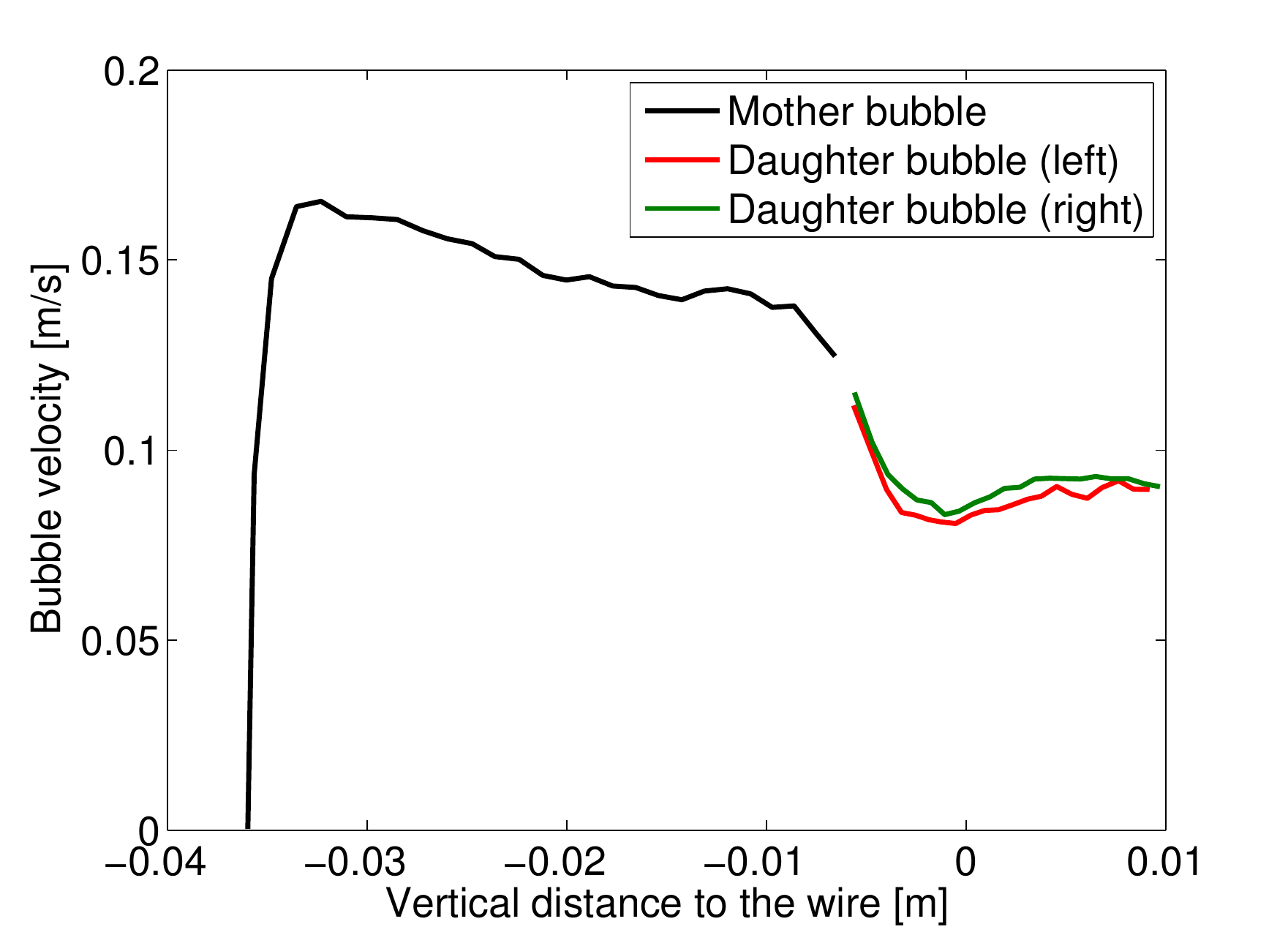}
 	} \subfloat[ $d_x = 0.5$]{
		\includegraphics[keepaspectratio=true, width=.48\textwidth]{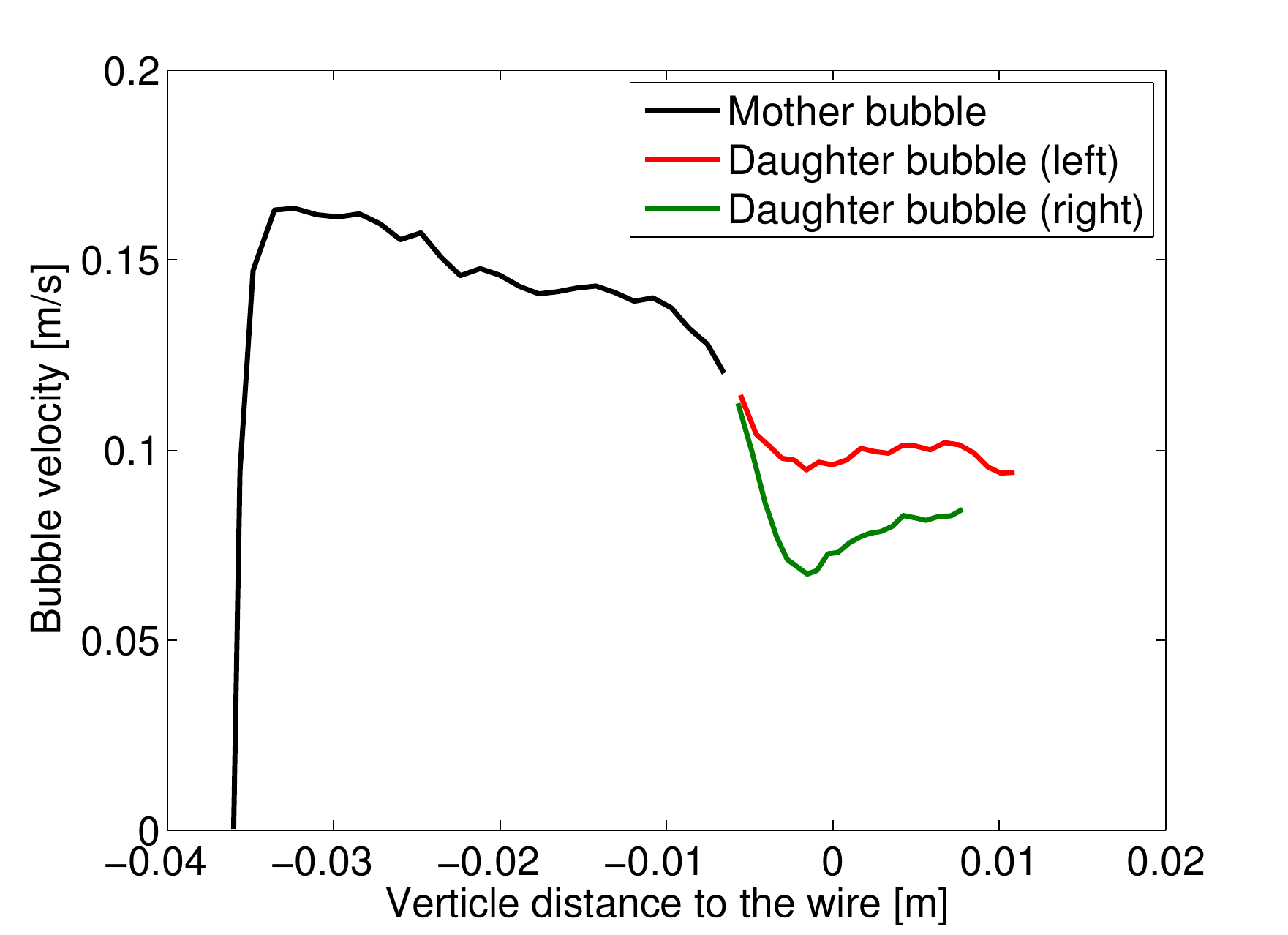}
	} 
	\captionsetup[subfigure]{margin=0pt} %parskip=0pt, hangindent=0pt, indention=0pt, singlelinecheck=true}
	\subfloat[$ d_x = 1 $]{
		\includegraphics[keepaspectratio=true, width=.48\textwidth]{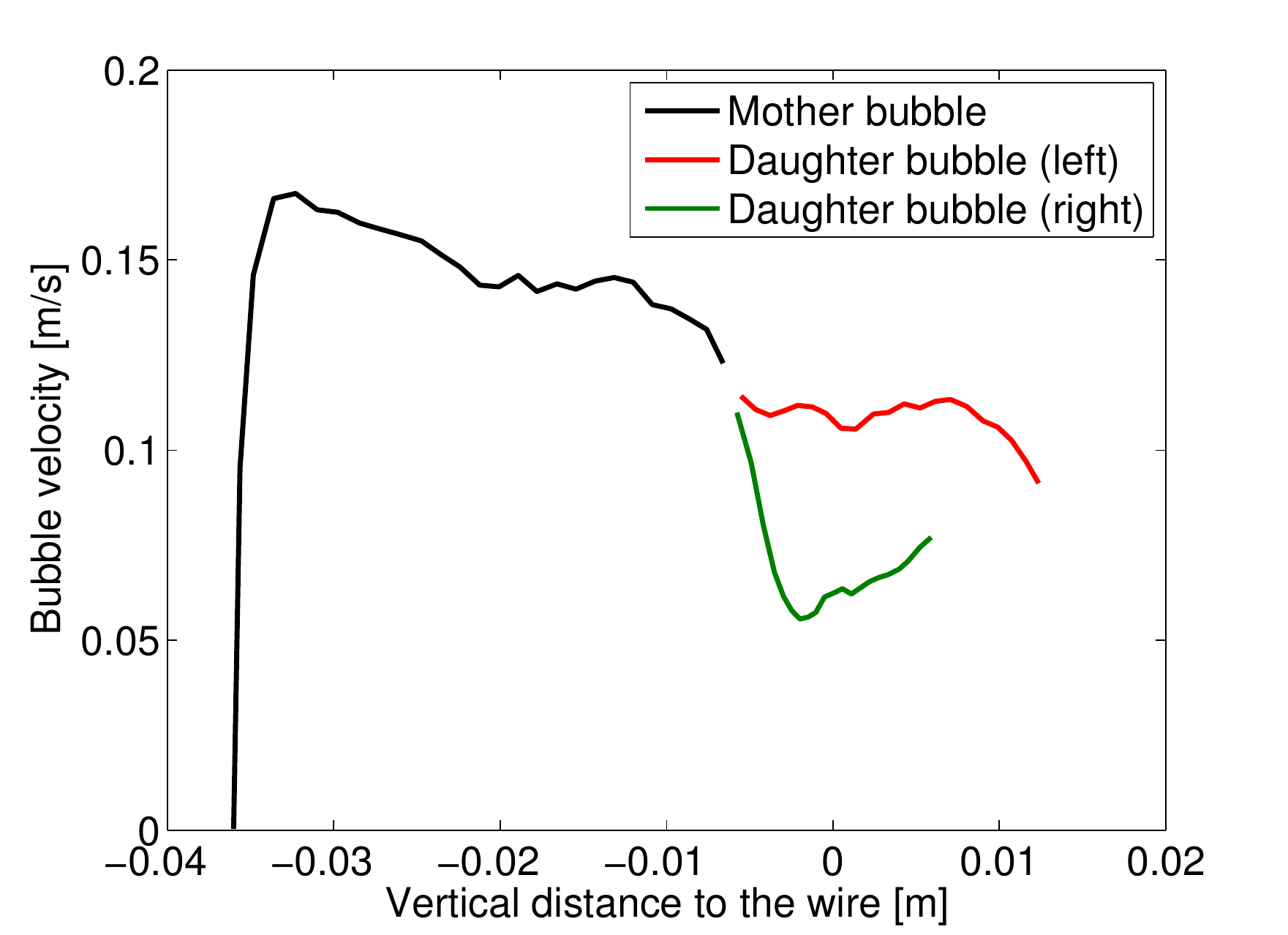}
 	} \subfloat[ $d_x = 1.5$]{
		\includegraphics[keepaspectratio=true, width=.48\textwidth]{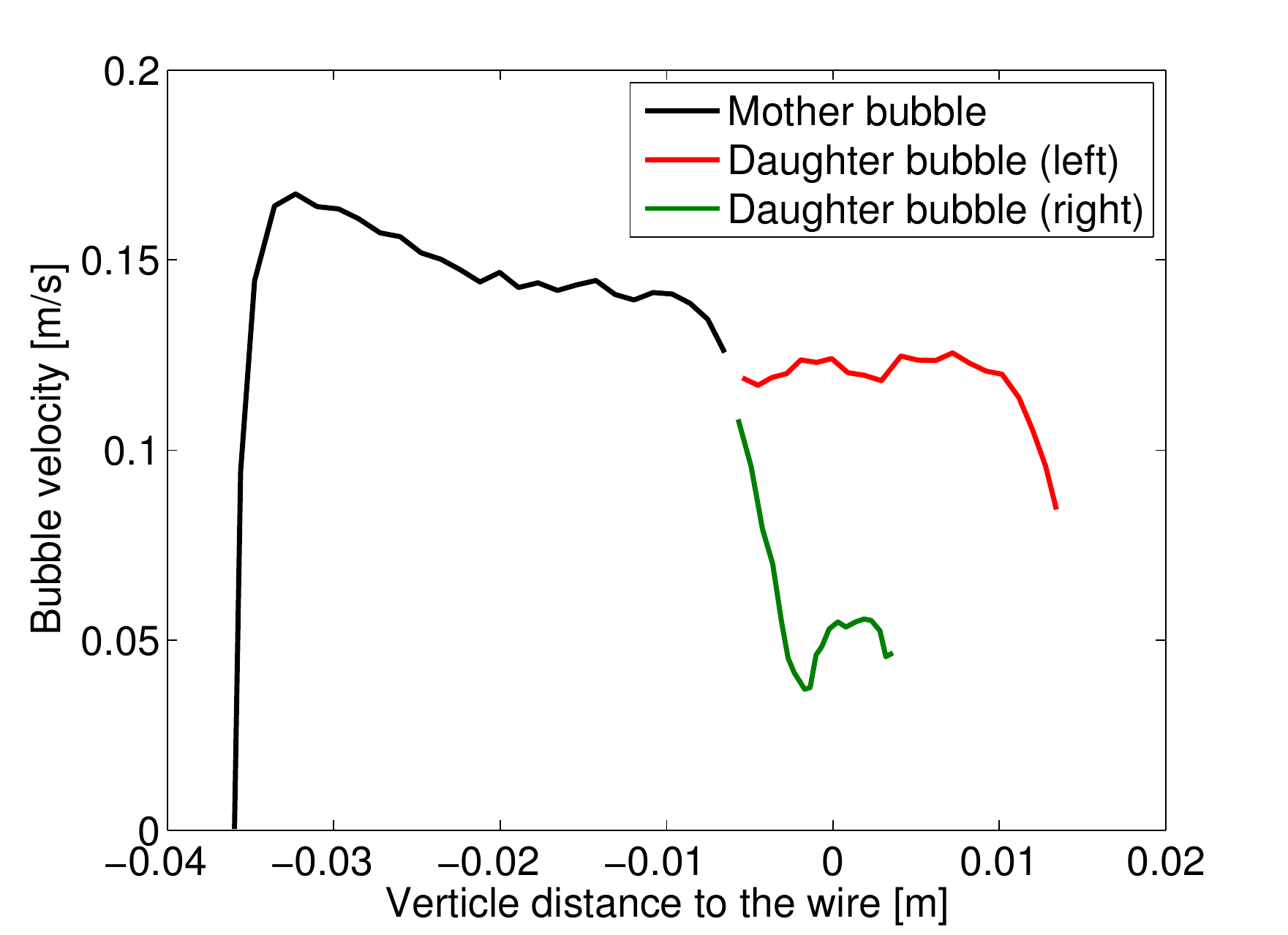} 
 	}   
 	\\
	\captionsetup{margin=20pt}
	\caption{ The rising velocities of the mother and the daughter bubbles. . }
	\label{cut_vel}
\end{figure}

\section{Concluding Remarks}
The cutting of bubbles at a single tube (wire) was investigated experimentally and numerically. For the simulations, a mesh free method was applied. The method enables a description of the deforming interface and the hydrodynamics of bubble cutting. For a first validation, we compared the solver to experimental data using the system air and water. A suffiently good agreement could be found in regard to bubble shape, bubble movement and cutting process itself. 
To study the effect of higher viscosity and  bubble position, a case study was done. One observes that the inital position of the bubble to the wire has a high impact on the final daughter bubble size ratio. A centric approach of the bubble to the wire leads to a cutting of the bubble in two equally sized daughter bubbles. By increasing the inital distance to the wire, the daughter-bubble size ratio  increases and the deviation between the velocities of daughter bubbles increases. Also, the movement of the bubbles directly behind the wire changes. While the bubbles split behind the wire at the centric approach, with increasing unsymmetry, the bubbles start to move inwards after the  initial  separation.
In future, further studies with overlapping wires are planned.
  
\section*{Acknowledgment}

This work is supported by the German research foundation, DFG grant KL 1105/27-1 and by RTG GrK 1932 “Stochastic Models for Innovations in the Engineering Sciences”, project area P1.


\begin{thebibliography}{30}

\bibitem{Mudde}
R. F. Mudde, and T. Saito
\textit{Hydrodynamical similarities between bubble column and bubbly pipe flow}, J. Fluid Mech. 437, 203–228, 2001.

\bibitem{Ahmed}
F. Shakir Ahmed, B. A. Sensenich , S. A. Gheni , D. Znerdstrovic, and M. H. Al Dahhan 
textit{Bubble Dynamics in 2D Bubble Column: Comparison between High-Speed Camera Imaging Analysis and 4-Point Optical Probe}, Chemical Engineering Communications 202, 85-95, 2014.

\bibitem{Choi}
K. H. Choi, and W. K. Lee
\textit{Comparison of Probe Methods for Measurement of Bubble Properties}, Chemical Engineering Communications 91, 35-47, 1990.


\bibitem{Prasser} 
H.-M.Prasser, D.Scholz, and C.Zippe
\textit{Bubble size measurement using wire-mesh sensors}, Flow Measurement and Instrumentation 12, 299-312, 2001.

\bibitem{BMJ} 
H.-J. Bart, M. Mickler, H.B. Jildeh 
\textit{Optical Image Analysis and Determination of Dispersed Multi Phase Flow for Simulation and Control}, In: Optical Imaging: Technology, Methods \& Applications, Akira Tanaka and Botan Nakamura (eds.), 1-63, Nova Science, N.Y. Lancaster, 2012.


\bibitem{Baltussen} 
M. W. Baltussen,
Bubbles on the cutting edge : direct numerical
simulations of gas-liquid-solid three-phase flows,
Preprint, TU Eindhoven, 2015. 

 

\bibitem{Chorin}
A. Chorin,
Numerical solution of the Navier-Stokes equations,
Math. Comput. vol. 22  (1968) 745-762. 


\bibitem{CSF}
J.U. Brackbill, D.B. Kothe, C. Zemach, 
\textit{A continuum method for modeling surface tension}, J. Comput. Phys. 100, 354�355, 1992.


\bibitem{Morris}
J. P. Morris, 
Simulating Surface Tension with Smoothed Particle Hydrodynamics, 
Int. J. Numer. Methods Fluids, 33 (2000) 333-353. 


%   
%\bibitem{eikonal} 
%A. Klar, S. Tiwari, and E. Raghavender, 
%\textit{Mesh Free method for Numerical Solution of The Eikonal Equation, Proceedings of International workshop on PDE Modelling and Computation},
%Advances in PDE Modelling and Computation, Ane Books Pvt. Ltd., 2013. 
 
\bibitem{DTKB}
C. Drumm, S. Tiwari, J. Kuhnert, H.-J. Bart,
\textit{Finite pointset method for simulation of the liquid�liquid flow field in an extractor}, Comput. Chem. Eng., 32, 2946, 2008. 

\bibitem{TKH16}
S. Tiwari, A. Klar, S. Hardt,
Numerical simulation of wetting phenomena by a meshfree particle method,
J. Comput. Appl. Maths., 292(216), 469-485. 

\bibitem{tiwari}
S. Tiwari, and J. Kuhnert,
\textit{Finite pointset method based on the projection method for simulations of the incompressible Navier-Stokes equations},
Meshfree Methods for Partial Differential Equations, eds. M. Griebel and M.A. Schweitzer, Lecture Notes in Computational Science and Engineering, Vol. 26 (Springer-Verlag, 2003), pp. 373-387.

\bibitem{TK07}
S. Tiwari, and J. Kuhnert,
\textit{Modelling of two-phase flow with surface tension by finite pointset method(FPM)},
J. Comp. Appl. Math, 203 (2007), pp. 376-386.
 

\end{thebibliography}
\end{document}